\documentclass{article}
\usepackage{emulateapj}
\usepackage{apjfonts}

\begin{document}

\newcommand{\figma}{1}
\newcommand{\figmb}{2}
\newcommand{\fignoise}{3}
\newcommand{\figmaschem}{4}
\newcommand{\figmbschem}{5}
\newcommand{\figmamap}{6}
\newcommand{\figmbmap}{7}
\newcommand{\figdist}{8}
\newcommand{\figlf}{9}
\newcommand{\figpagblf}{10}
\newcommand{\figmacmd}{11}
\newcommand{\figmbcmd}{12}
\newcommand{\figms}{13}
\newcommand{\figehblf}{14}
\newcommand{\figsim}{15}
\newcommand{\figdens}{16}
\newcommand{\figmaspec}{17}
\newcommand{\figmbspec}{18}
\newcommand{\figext}{19}

\submitted{To be published in The Astrophysical Journal}
\title{Color--Luminosity Relations for the Resolved
Hot Stellar Populations in the Centers of M~31 and M~32$^\dagger$}

\author{Thomas M. Brown}

\affil{Laboratory for Astronomy \& Solar Physics, Code 681, NASA/GSFC\\
Greenbelt, MD 20771\\tbrown@band1.gsfc.nasa.gov}

\smallskip

\author{Henry C. Ferguson}
\affil{Space Telescope Science Institute\\
3700 San Martin Drive, Baltimore, MD 21218\\ferguson@stsci.edu}

\smallskip

\author{S. A. Stanford}
\affil{Institute of Geophysics and Planetary Physics, Lawrence Livermore
National Laboratory\\
Livermore, CA 94550\\ adam@igpp.llnl.gov}

\and

\author{Jean-Michel Deharveng}
\affil{Laboratoire d'Astronomie Spatiale du CNRS, Traverse du Siphon, BP 8\\
F-13376 Marseille Cedex 12, France\\ jmd@astrsp-mrs.fr}

\begin{abstract}

We present {\it Faint Object Camera (FOC)} ultraviolet images of the central
$14 \times 14 \arcsec$
of Messier 31 and Messier 32.  The hot stellar population detected
in the composite UV spectra of these nearby galaxies is partially resolved
into individual stars, and their individual colors and apparent magnitudes are
measured.  We detect 433 stars in M~31 and 138 stars in M~32, 
down to detection
limits of $m_{F275W}$~=~25.5~mag and $m_{F175W}$~=~24.5~mag.
We investigate
the luminosity functions of the sources, their spatial distribution,
their color-magnitude diagrams, and their total integrated far-UV flux.
Comparison to {\it IUE} and {\it HUT} spectro-photometry and {\it WFPC2} 
stellar photometry indicates consistency at the 0.3~mag level, with possible
systematic offsets in the {\it FOC} photometry at a level less than this. 
Further calibrations or observations with 
the {\it Space Telescope Imaging Spectrograph (STIS)} will be necessary to 
resolve the discrepancies.  Our interpretation rests on the assumption that
the published {\it FOC} on-orbit calibration is correct.

Although M~32 has a weaker UV upturn than M~31, 
the luminosity functions and color-magnitude diagrams 
of M~31 and M~32 are surprisingly similar,
and are {\it inconsistent}
with a majority contribution from any of the following: PAGB stars
more massive than 0.56~$M_\odot$ (with or
without associated planetary nebulae), main 
sequence stars, or blue stragglers.

Both the luminosity functions and color-magnitude diagrams are 
{\it consistent} with a dominant population of stars that have evolved
from the extreme horizontal branch (EHB) along tracks with masses
between 0.47 and 0.53 $M_\odot$.  These stars are well below the detection
limits of our images while on the zero-age EHB, but become detectable 
while in the more luminous (but shorter) AGB-Manqu$\acute{\rm e}$ 
and post-early
asymptotic giant branch (PEAGB) phases.  The {\it FOC} observations require
that only a very small fraction of the main sequence population 
(2\% in M~31 and 0.5\% in M~32) in these two galaxies 
evolve though the EHB and 
post-EHB phases, with the remainder evolving
through bright PAGB evolution that is so rapid that few if any stars are 
expected in the small field of view covered by the {\it FOC}.
A model with a flat EHB star mass distribution
reproduces the {\it HUT} and {\it IUE} spectra of
these two galaxies reasonably well, although there is some indication
that an additional population of very hot 
(T$_{\rm eff} > 25000$ K) EHB stars may 
be needed to reproduce the {\it HUT} spectrum of M~31 near the Lyman limit,
and to bring integrated far-UV fluxes of M~31 and M~32 into agreement
with {\it IUE}. 

In addition to the post-EHB population detected in the {\it FOC}, we find
a minority population ($\sim$ 10\%) of brighter stars that populate a region
of the CMD that cannot be explained by canonical post-HB evolutionary tracks.
The nature of these stars remains open to interpretation.

The spatial distributions of the resolved UV-bright stars in both galaxies 
are more centrally concentrated than the underlying diffuse emission,
implying that stellar populations of different age and/or metallicity might 
be responsible for each component.

\end{abstract}


\noindent $\dagger$
Based on observations with the NASA/ESA Hubble Space Telescope obtained
at the Space Telescope Science Institute, which is operated by the
Association of Universities for Research in Astronomy, Incorporated,
under NASA contract NAS~5-26555.

\newpage

\section{INTRODUCTION}

The spectra of elliptical galaxies and spiral galaxy bulges exhibit
a strong upturn shortward of 2700~\.{A}, dubbed the ``UV upturn.''  
At the time of its discovery, the 
existence of a hot stellar component went against
the traditional picture of early-type galaxies.
The canonical view of ellipticals
held that these galaxies contained a cool, passively evolving population
of old stars.  The pioneering UV observations of ellipticals -- with
the {\it Orbiting Astronomical Observatory (OAO)} (Code \& Welch 
1979\markcite{CW79})
and the {\it International Ultraviolet Explorer (IUE)} 
(Bertola, Capaccioli, \& Oke 1982\markcite{BCO82}) -- 
could only sample the Rayleigh-Jeans tail of the hot UV flux, with
poor signal-to-noise and resolution.  
Early explanations for the source of the UV upturn 
covered a wide range of candidates, including massive young stars,
hot horizontal branch stars, planetary nebula nuclei, and several
binary scenarios (see Greggio \& Renzini 1990\markcite{GR90}
for a complete review).
The presence of young stars would imply ongoing star formation in 
early-type galaxies, while the evolved candidates suggested that
old stellar populations could be efficient UV emitters.  

Characterized by the $m_{1550}-V$ color, the
UV upturn shows surprisingly strong variation (ranging from 2.05--4.50 mag) in 
nearby quiescent early-type galaxies 
(Bertola et al.\ 1982\markcite{BCO82};
Burstein et al.\ 1988\markcite{B88}), even though
the spectra of ellipticals at longer wavelengths
are qualitatively very similar.  A large sample of UV measurements
demonstrated that the UV upturn
is positively correlated with the strength of Mg$_2$ line absorption
in the $V$ band, in the sense that the $m_{1550}-V$ color is bluer at
higher line strengths, opposite to the behavior of optical color indices
(Burstein et al.\ 1988\markcite{B88}).  
Opposing theories have been devised to explain this correlation.
For example, Lee (1994\markcite{L94}) and Park \& Lee (1997\markcite{PL97})
suggest that the UV flux originates in the
low metallicity tail of an evolved stellar population with
a wide metallicity distribution.  Their reasoning is that the more
massive ellipticals formed earlier, and that it is actually the 
{\it mean} metallicity that is
higher (and driving the optical indices) in the older and bluer galaxies.
In contrast, several groups
(Brocato et al.\ 1990\markcite{BMMT90};
Bressan, Chiosi, \& Fagotto 1994\markcite{BCF94};
Greggio \& Renzini 1990\markcite{GR90};
Horch, Demarque, \& Pinsonneault 1992\markcite{HDP92};
Brown et al.\ 1997\markcite{B97}; Yi, Demarque, 
\& Oemler 1997\markcite{YDO97}) argue that metal-rich
horizontal branch stars and their progeny are responsible for the UV
flux.  Under the metal-rich hypothesis, the canonical trend in horizontal
branch morphology (i.e., redder HBs with higher metallicity) is reversed
at high metallicity, due to increased helium abundance and
possibly a higher mass loss rate on the red giant branch, resulting
in the production of hot UV-efficient stars on the extreme horizontal branch
(EHB).  These hypotheses lead to different
ages for the stellar populations in these galaxies.  Ages exceeding those
of Galactic globular clusters are required under the
low-metallicity Park \& Lee
(1997\markcite{PL97}) hypothesis, while ages as low as 8 Gyr are allowed in 
the Bressan et al.\ (1994\markcite{BCF94}) model.
In these two scenarios, the EHB
stars are drawn from either 
tail of the metallicity distribution.  However,
it is also possible, indeed perhaps more likely, that the EHB stars
arise from progenitors near the peak of the metallicity distribution
(cf.\ Dorman, O'Connell, \& Rood 1995\markcite{DOR95}),
but represent a relatively rare occurrence.  
The correlation of $m_{1550}-V$ with the global metallicity of the galaxy
might indicate that this rare path of stellar evolution becomes less so
at high metallicity and helium abundance.

The {\it Hopkins Ultraviolet Telescope (HUT)}, designed for
medium resolution ($\approx$~3~\.{A}) spectroscopic observations
of faint extended UV sources down to the Lyman limit at 912~\.{A},
offered a new perspective on these populations.
With observations of two galaxies on
the Astro-1 mission (M~31 and NGC~1399), Ferguson 
et al.\ (1991\markcite{F91}) 
demonstrated that young, massive stars cannot be a 
significant contributor to the UV upturn.  There is a lack of strong
\ion{C}{4} absorption expected from such stars, and the continuum flux
decreases from 1050~\.{A} down to the Lyman limit.  Such a decrease
implies that the UV flux is dominated by stars with temperatures
$\leq 25000$~K and is incompatible with emission by a population of
young stars having a normal initial mass function.  
Ferguson et al.\ (1991\markcite{F91}) also 
suggested that a bimodal distribution on the horizontal
branch was needed to reproduce the shape of spectra from the near-UV to
the far-UV, otherwise the spectra would be flatter than observed.
Six more galaxies
were observed on the Astro-2 mission (M~49, M~60, M~87, M~89, NGC~3115, and
NGC~3379), and with these data Brown et 
al.\ (1997\markcite{B97}) demonstrated that
a two-component population of high-metallicity post-HB stars could
reproduce the UV light seen in nearby ellipticals.  In this model,
most ($> 80$\%) of the UV-producing stars were undergoing 
post-asymptotic-giant-branch (PAGB) evolution,
and the remainder were evolving along AGB-Manqu$\acute{\rm e}$ paths from the
extreme horizontal branch.  Although in the minority, these 
AGB-Manqu$\acute{\rm e}$ stars can produce the majority of the flux, because
their lifetimes are orders of magnitude longer than those of the PAGB
stars.

Because even the brightest of these galaxies are faint and extended
in the UV, studies of the UV upturn have focused mostly on
the composite spectral energy distributions of ellipticals.
However, the UV imaging capabilities of {\it HST} 
have now opened the possibility
of studying the resolved UV population, at least in the nearest
galaxies. Attempts to do this prior to the {\it HST} 
refurbishment were undertaken
by King et al.\ (1992\markcite{K92}), for M~31, and 
by Bertola et al.\ (1995\markcite{BBB95}), for M~31 and M~32.

King et al.\ (1992\markcite{K92}) obtained a pre-COSTAR {\it Faint
Object Camera (FOC)} observation of a $44 \times 44 \arcsec$ field in
the center of M~31, using the F175W filter and the F/48 relay.  
They found more than 100
sources that they identified as PAGB stars.  Intermediate-mass ($M >
0.6~M_{\odot}$) PAGB stars are short-lived, and given a population size
constrained by the fuel consumption theorem, the large number of
detected stars implied that these were low-mass PAGB stars.  King et
al.\ (1992\markcite{K92}) estimated that these stars accounted for
approximately 20\% of the UV light, with the rest unresolved,
presumably coming from EHB stars and their AGB-Manqu$\acute{\rm e}$ 
descendants, which could account for
this unresolved light without violating fuel consumption constraints.

Bertola et al.\ (1995\markcite{BBB95}) used the {\it FOC} to image
M~31, M~32, and NGC~205 with the combination of the F150W and F130LP
filters on the F/48 relay.  
Although the optical luminosity enclosed by the {\it FOC}
field was higher in M~32 than in M~31, they found far fewer UV sources
in the M~32 field.  Because M~32 has a weaker UV upturn and lower
metallicity, the UV light was expected to originate in PAGB stars of
higher mass (and shorter lifetimes) than those in M~31; so, the
relative numbers of detected sources were in line with these
expectations.  However, the luminosity functions in M~31 and M~32
appeared similar, in contrast to expectations when comparing a
population of less massive PAGB stars to a population of more massive
ones.  Such a puzzle may be partly explained if the PAGB stars are
enshrouded in dust during the early part of their evolution away from
the AGB.

Both previous {\it FOC} studies faced daunting challenges in untangling the
uncertainties in the {\it FOC} sensitivity calibration and the red leak of the
filters.  King et al.\ (1992\markcite{K92}) adopted the best 
in-flight calibration at the time, and used ground calibrations of
the filter and photocathode response to assess the effects of red leak.
They also determined that the 
pre-COSTAR PSF required a huge aperture correction of 2.6~mag. 
Bertola et al.\  (1995\markcite{BBB95}) 
derived an independent calibration based upon {\it IUE} observations
of NGC~205, M~31, and M~32.  In the Bertola et al.\ calibration, 
the nominal {\it FOC} F150W+F130LP efficiency curve required multiplication
by factors of 0.21--0.92 (varying with wavelength) 
in order to produce agreement between the {\it FOC} and {\it IUE}.
Checking their revised calibration against common stars in the King
et al.\ (1992\markcite{K92}) F175W images, 
Bertola et al.\ determined that the F175W
efficiency curve also required revision, such that the peak in the efficiency
curve was at 28\% of its nominal value, but with increased red leak from
longer wavelengths.  We demonstrate in \S\ref{secf48} that the pre-COSTAR
{\it FOC} data calibration was seriously in error.

We have used the 
refurbished, recalibrated {\it FOC} to follow up
these earlier studies with deeper UV images of the M~31 and M~32 cores,
in order to further characterize the evolved stellar populations in
these galaxies.  Our observations use the F175W and F275W filters
to determine color-luminosity relationships in these galaxies, and to
compare them to the predictions of stellar evolutionary theories.

\section{OBSERVATIONS}
\label{secobs}

In February of 1994 and 1995, we obtained 
{\it Faint Object Camera (FOC)} 
ultraviolet images of the giant spiral galaxy Messier 31 (M~31, NGC~224) 
and its compact elliptical companion, Messier 32 (M~32, NGC~221).
In comparison to the other galaxies with measured UV upturn strengths,
M~31 and M~32 show relatively weak upturns,
with respective $m_{1550}-V$ colors of 3.51 and 4.50 mag
(Burstein et al.\ 1988\markcite{B88}).  
Our images are shown in Figures \figma\ and \figmb\ (Plates XX and YY), 
as combined with archival 
{\it Wide Field Planetary Camera 2 (WFPC2)} data in the optical
(F555W) filter.  Although the images each include the background from the
bright galaxy core, we can clearly resolve many of the hot post-horizontal 
branch (post-HB) stars that are responsible for the UV light.
Our observations were with the 
{\it FOC} F/96 relay, using the $512 \times 1024$ zoomed format 
($512z \times 1024$); this format provided 
the full $14 \times 14\arcsec$ field of view at the expense of the full
dynamic range available in the {\it FOC}.   Thus, after processing via
the standard pipeline, the frames required dewrapping of brightly illuminated
pixels.  Once dewrapped, the frames from a given
band were aligned and coadded.  The observations are summarized in 
Table~\ref{tabobs}.

Because the standard pipeline processing applies a dezooming of the
pixels (from $512 \times 1024$ to $1024 \times 1024$) and a geometric
correction, the signal in neighboring pixels is correlated, and so
the noise characteristics of these images significantly
deviate from the Poisson distribution.  Well-defined source detection
required modelling of the noise characteristics in the images, especially
given the presence of a nonuniform background from the underlying galaxies.
To simulate the {\it FOC} noise characteristics, we applied the same dezooming
and geometric correction to a set of uniform frames that had
Poisson noise; this set of test
frames covered the range of signal appropriate
to our galaxy images.  
We then fit a second-order polynomial to the pixel-to-pixel variance in
each simulation; the fit is shown in Figure~\fignoise.   
Poisson statistics
significantly overestimate the noise in this format up to a signal of
approximately 500 counts per pixel, and underestimate the noise
beyond this level.

We fit the full-width half maximum (FWHM) of the point spread function (PSF)
in each image using a selection of isolated stars.  The fits were
performed with the IRAF routine FITPSF, and the results are 
listed in Table~\ref{tabobs}.
It appears that the focus was slightly better in the M~32 images,
perhaps due to ``breathing'' in the
telescope structure.  The sharpness of the PSF is known to improve
at longer wavelengths, so it is not surprising that the F175W images
show wider PSFs as compared to the F275W images.  We note that
the PSF does not appear to vary significantly with position in the {\it FOC}
field.  Our fits to the PSF were
used to determine source detection criteria, to simulate completeness,
and to simulate spurious source detections, all of which are discussed below.
None of the sources appear significantly resolved (e.g., as planetary nebulae),
although two sources near the top of the M~31 field appear somewhat elongated
in both the F175W and F275W images.  These two stars have been flagged in our 
catalog (Table~\ref{tabm31cat}), and are discussed further in \S\ref{secphot}.
One does not expect to resolve planetary nebulae (PNe)
at a distance of 770 kpc.  PNe studies (e.g., Schneider \& Buckley
1996\markcite{SB96}) show that very few PNe have diameters as large as
0.8 pc, with most smaller than 0.2 pc.  At the M~31 distance, a PN with
a diameter of 0.2 pc would span four {\it FOC} pixels.

The in-flight calibration of the {\it FOC} F175W and F275W filters 
was done using observations of HZ4 in the $256 \times 256$ format, 
in conjunction with neutral 
density filters due to count rate limitations
(Jedrzejewski 1996\markcite{J96}). Thus, our exact combination
of format and filters has not been calibrated directly, and this may contribute
to some of the systematic uncertainties discussed in \S\ref{secf275}.

While the primary purpose of our observations was to study the
stellar population, at the same time these new {\it FOC} images provide
the most detailed view of the morphology of the double nucleus of M~31.
{\it Planetary Camera (PC)} observations by
Lauer et al.\ (1993\markcite{L93}) provided the first clear
optical view of the double nucleus,
and deconvolved {\it FOC} UV observations (King, Stanford, \&
Crane 1995\markcite{K95}) indicated that the twin peaks are not
due to a dust lane in the center of the galaxy.  Our own images of M~31
(Plate XX) clearly show the nuclei as separate entities, also without
any evidence of a dust lane.  
The UV-bright nucleus manifests itself as a sharp
peak offset from the optically-bright nucleus.  The peak is clearly
not a point source, but appears to be a tight cluster of UV bright 
stars.  The color of the UV-bright nucleus is 
$m_{F175W} - m_{F275W} = -0.34$~mag
within 0.2$\arcsec$, consistent  with stars of effective temperature 11500~K.
For a discussion of the M~31 nuclei
and their respective colors, see King et al.\ (1995\markcite{K95}) and
references therein.

We note that the small field of view does not include any 
previously published planetary nebulae or globular clusters in the inner
regions of M~31 and M~32, according to a search through the SIMBAD database.
Observations of these objects are difficult so close to the core of a galaxy.

\section{DATA REDUCTION}

\subsection{Source Detection}
\label{secdet}

Source detection and photometry in the {\it FOC} images are complicated by
the strongly varying background.  Most existing photometry packages,
even sophisticated ones, require the user to input a uniform counts
threshold above which a signal is considered a detection.  Setting this
threshold at a high level, 

\begin{figure*}
\parbox{7.0in}{\epsfxsize=7.0in \epsfbox{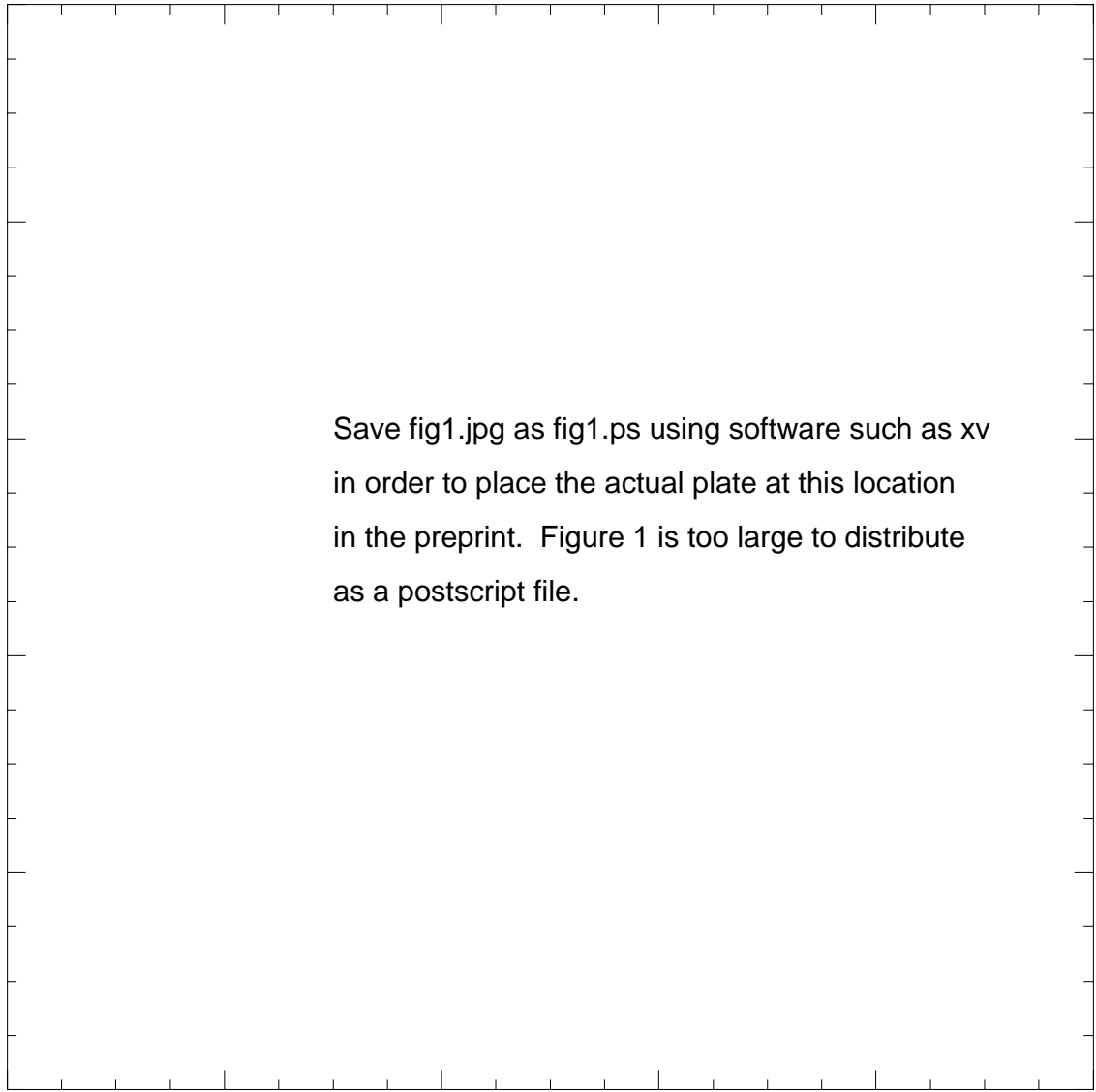}}
\caption{
This 3-color image of the M~31 center was created by assigning 
the {\it FOC} F175W
image to the blue, the {\it FOC} F275W image to the green, and an archival
{\it WFPC2} F555W image (ID\# 5464) to the red.  The double nucleus is clearly 
distinguished, with one nucleus brighter in the visible and the other 
brighter in the UV.  The two nuclei are clearly not the result of a dust
lane.  The hot post-HB stars densely populate the field.
We note that the F555W image is saturated in the core.
}
\end{figure*}

\clearpage

\begin{figure*}
\parbox{7.0in}{\epsfxsize=7.0in \epsfbox{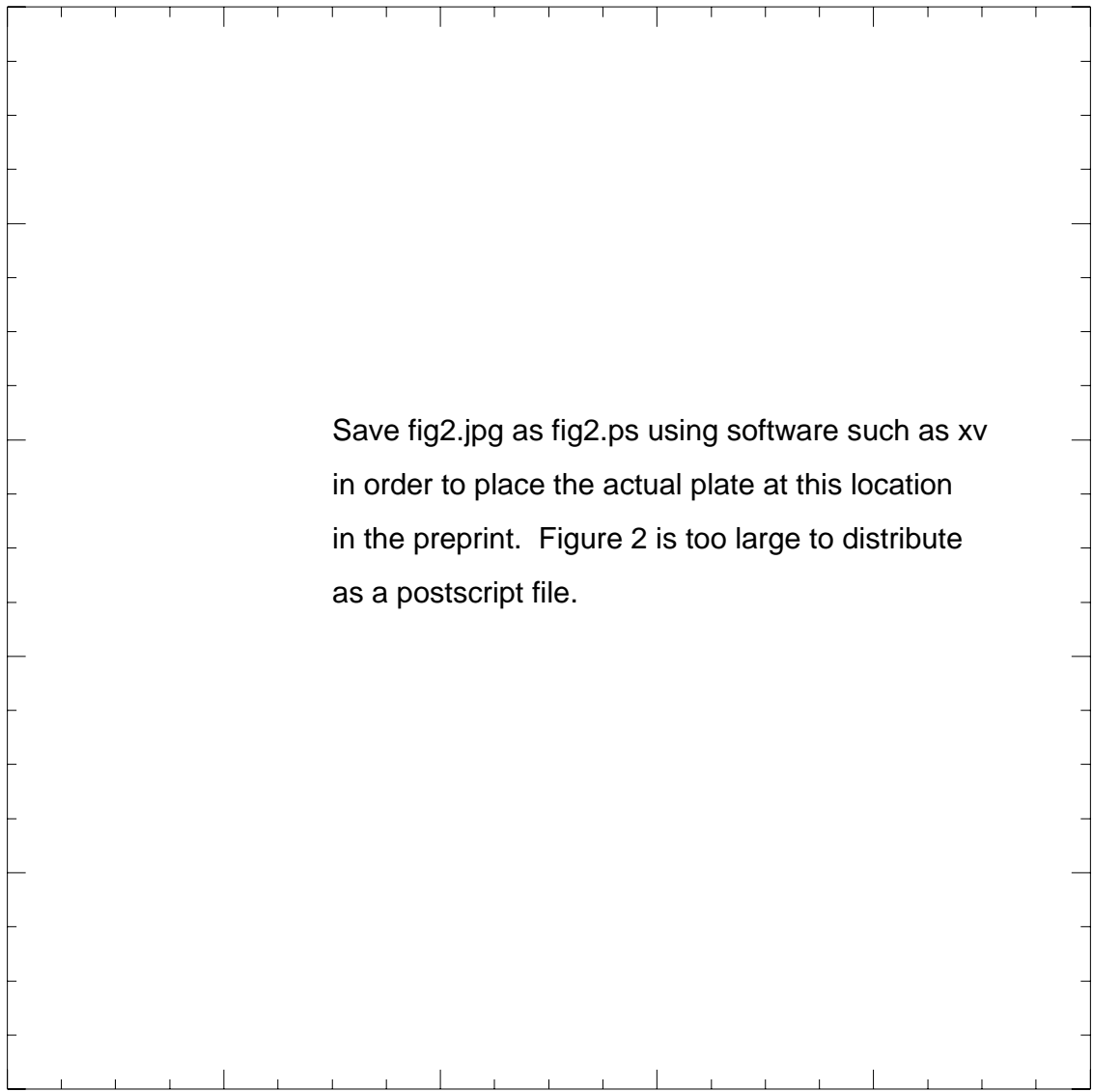}}
\caption{
The center of M~32, using the same filters as
Figure~\figma\ (the F555W archival
image again has ID\# 5464).  
In comparison to M~31, the center of M~32 is much smoother
and shows far fewer stars.  As in the previous figure,
the F555W image is saturated in the core.
}
\end{figure*}

\clearpage

\noindent
appropriate to the noise characteristics in
the brighter regions of an image, will cause the algorithm to miss
fainter sources that should be detectable in the low-background regions
of the image.  Setting the threshold at a low level, appropriate to the
noise in the fainter regions of an image, will cause the algorithm to
find many spurious sources in the bright-background regions.

\parbox{3.0in}{\epsfxsize=3.0in \epsfbox{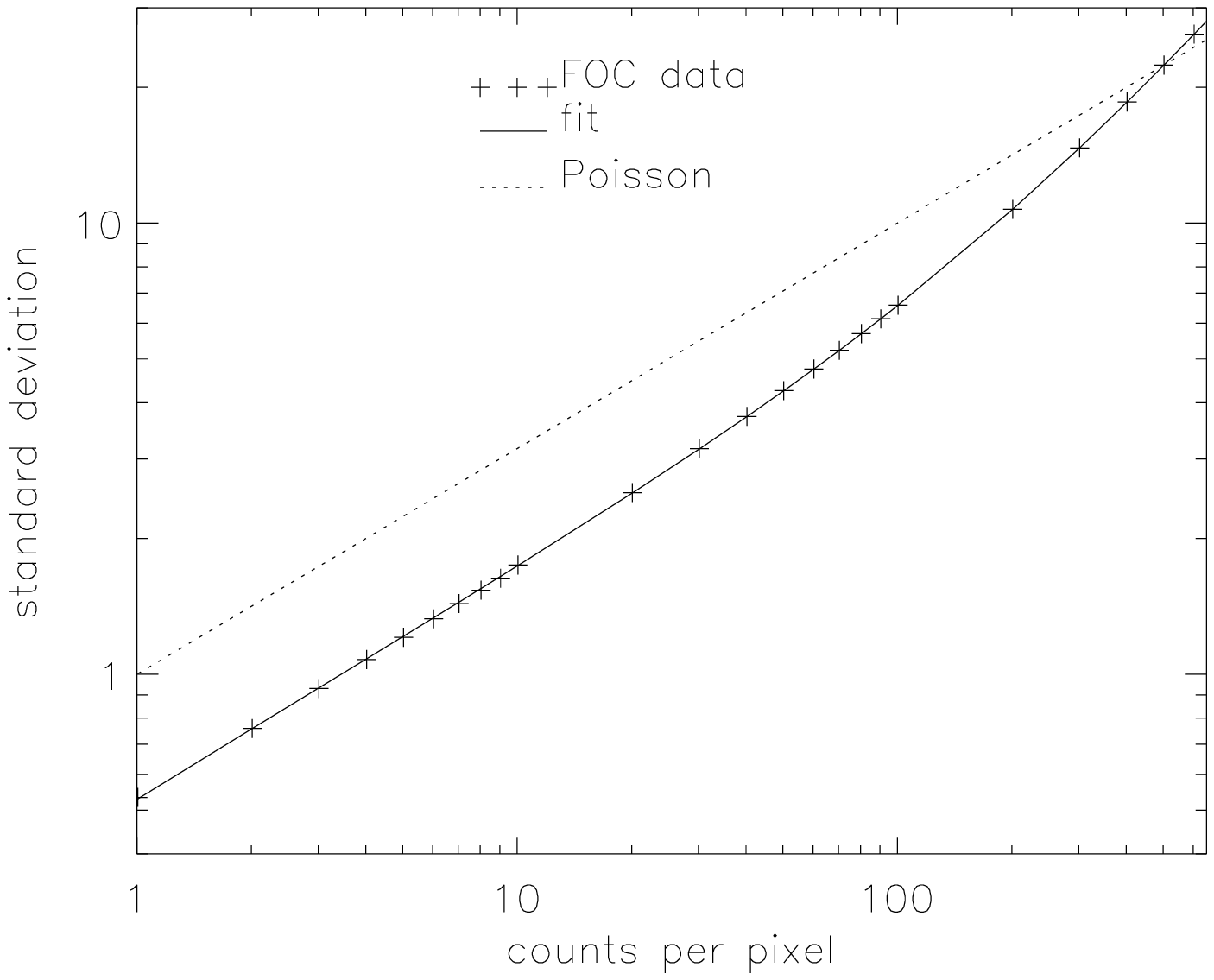}}
\centerline{\parbox{3.0in}{\small {\sc Fig.~\fignoise--}
The noise characteristics of the {\it FOC} data (crosses)
in the F/96 $512z \times 1024$ format deviate significantly
from the Poisson distribution (dotted).  Our polynomial fit to the
{\it FOC} standard 
deviation vs. counts (solid) is used to set a spatially constant
threshold for source detection in the nonuniform galaxy images (see text).
}}\vspace{0.2in}
\addtocounter{figure}{1}

To work around this difficulty, we produced a galaxy-free, constant-noise
image for each band in both galaxies, and used these ``detection'' images 
for source detection.  The detection images were created by taking the
original images, applying a median filter ($35 \times 35$ pixels), and 
subtracting the median image from the original.  The result was
a ``galaxy-free'' image, but not one with constant noise characteristics.
Next, we used our polynomial fit for the variance (Figure~\fignoise) to 
convert the median image into a ``noise'' image, and divided the galaxy-free
image by this noise image.  The result was a galaxy-free image where
the standard deviation in the background was unity across the entire image.
We then applied the IRAF task DAOFIND, provided under the DAO
Crowded-Field Photometry Package (DAOPHOTX), setting our detection limit
at 4 $\sigma$.  The package looks for local density maxima
that meet user-supplied criteria for FWHM, roundness, and
sharpness.  We left the criteria for roundness and sharpness at the default
settings, but after some experimentation, we set the 
FWHM to 5 pixels for each band.  Although the
PSF seems to vary somewhat from image to image (see \S\ref{secobs} and
Table~\ref{tabobs}), we wanted to be able to detect any nebulae which might not
have the narrow PSF of a star.  
While this procedure is effective at pushing the source 
detections to the noise limits of the observations, it has
the consequence that the limiting magnitude is a function of
position.  For most of the subsequent discussion, we limit ourselves
to the brighter sources and to a region of each galaxy more than
1.5$\arcsec$ from the nucleus.

DAOFIND produced source lists for each band.  Due to the geometric correction,
the edges of the {\it FOC} detector do not coincide with the edges of the image
(see Figures~\figma\ \& \figmb).  This causes problems with the 
source detection in the images at the detector boundaries, necessitating the  
rejection of sources along the edges of the detector 
(see Figures~\figmaschem\ and \figmbschem).
We combined the source lists for each band into a master source list for each
galaxy, but retained the information about the band(s) in which
each source was detected.  Thus, a star that was too faint to be a 
4~$\sigma$ detection in the F175W filter but bright in the F275W filter 
could still produce F175W photometry, and vice versa.
These matters complicate the characterization of our completeness vs. 
magnitude, as we discuss in \S\ref{seccompspur}.
Sources detected in both bands were matched by taking
the position in one band and finding the closest detection in the other band
within 3 pixels.

\subsection{Photometry}
\label{secphot}

We used the STMAG system for our analysis of the {\it FOC} data.  In this 
system, the image header parameter PHOTFLAM defines the inverse sensitivity.
The quantities of interest are:\\
\[m = {\rm -2.5 \times log_{10}} f_{\lambda} -21.10\]
\[f_{\lambda} ={\rm counts \times PHOTFLAM / EXPTIME}\]
\[m = {\rm -2.5 \times log_{10} (counts/EXPTIME)} + m_o\]
\noindent
where EXPTIME is the exposure time, $m_o$ is the magnitude that produces
1 count per second in the observing mode, and -21.10 is the magnitude
corresponding to a flat flux 
($f_{\lambda}$) of 1 erg s$^{-1}$ cm$^2$ \.{A}$^{-1}$
in the bandpass.  The values of PHOTFLAM, EXPTIME, and $m_o$ for our data 
are given in Table~\ref{tabobs}.

For our aperture photometry, we used the IRAF task PHOT, 
also provided under the DAOPHOTX package.  In each band for each galaxy,
we used a circular aperture with a radius of 3 pixels (0.042$\arcsec$),
centered on the coordinates found during detection.
If the source was detected in both bands, the coordinates
used for the photometry in a given band matched those determined during 
detection in that band.  If the source was detected in only one band,
then those coordinates were used for the photometry in both bands.
We did not recenter during the photometry, because a star might be
very faint in one band while bright in the other, possibly resulting in an 
erroneous shift to a nearby local maximum in the faint band photometry. 
Sky subtraction was performed by taking an annulus with an inner radius
of 15 pixels and an outer radius of 20 pixels; the background level
was taken to be the centroid of the histogram of pixel intensities
within that annulus.  By applying varying apertures to isolated stars, 
we found that an aperture correction of -0.56 mag was needed for the 
F275W photometry and an aperture correction of -0.61 mag was needed for 
the F175W photometry.  

Our photometric catalogs were edited to reject stars within 1.5$\arcsec$ 
of the center in each galaxy (see Figures~\figmaschem\ and 
\figmbschem), because the spurious sources 
found near the cores would be very bright (see \S\ref{seccompspur}).
Furthermore, because a star could be detected in one
band but not the other, if the photometry failed in either band for
a star, it was also rejected.  Failed photometry occurred if the counts
in the 3-pixel aperture were not positive after subtraction of the sky
background.  Our source catalogs are shown in Tables~\ref{tabm31cat} 
and \ref{tabm32cat}, truncated at the detection limits of 
$m_{F175W} = 24.5$~mag and $m_{F275W} = 25.5$~mag (see \S\ref{seccompspur}).
We also plot maps of these catalogs in Figures~\figmamap\ and 
\figmbmap, to show the distribution of these stars in our fields.
Note that in Tables~\ref{tabm31cat} and \ref{tabm32cat}, 
a ``:'' in a magnitude denotes a measurement on a star that was not
originally detected in that band.

The two brightest stars in our photometric catalog for M~31 have been
flagged in Table~\ref{tabm31cat} because of photometric problems
(the brightest is marked ``a'' and the next brightest ``b'').
For consistency, their values in the table are the result of our standard
reduction procedure, but there are two problems that would change these 
values.  First,
the stars appear somewhat elongated in both {\it FOC} bands, and thus
the aperture corrections

\newpage

\parbox{5.75in}{\epsfxsize=5.75in \epsfbox{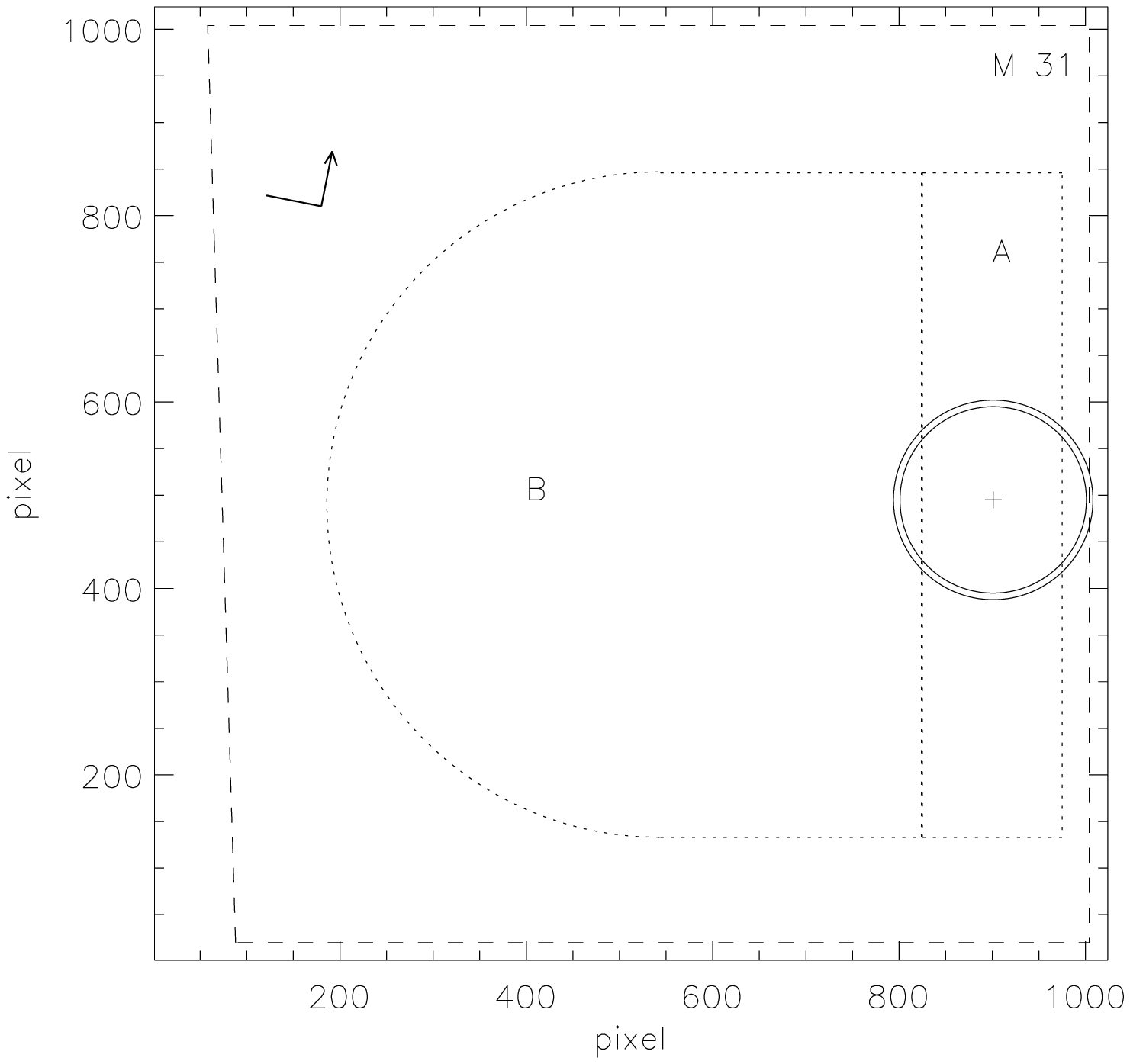}}
\vskip-3.0in
\hspace*{5.0in}
\parbox{2.0in}{
\small {\sc Fig.~\figmaschem--}
A schematic of our M~31 analysis clarifies the complicated procedures
described in the text.  We reject sources near and beyond the edge of the
detector (dashed) because the source detection algorithm is confused by
the abrupt change at the detector boundary.  We also reject sources detected
within 1.5$\arcsec$ (larger circle) of the core (cross).  The flux within
1.4$\arcsec$ (smaller circle) of the core is used to investigate any 
systematic errors in the {\it FOC} calibration.  
Regions ``A'' (dotted rectangle centered on core)
and ``B'' (dotted semicircle and rectangular region) can be used to
approximate the $10 \times 20\arcsec$ oval {\it IUE} aperture, by taking
A+2B.  North and East on the sky are shown by the compass.
}
\vspace*{1.25in}
\addtocounter{figure}{1}

\parbox{5.75in}{\epsfxsize=5.75in \epsfbox{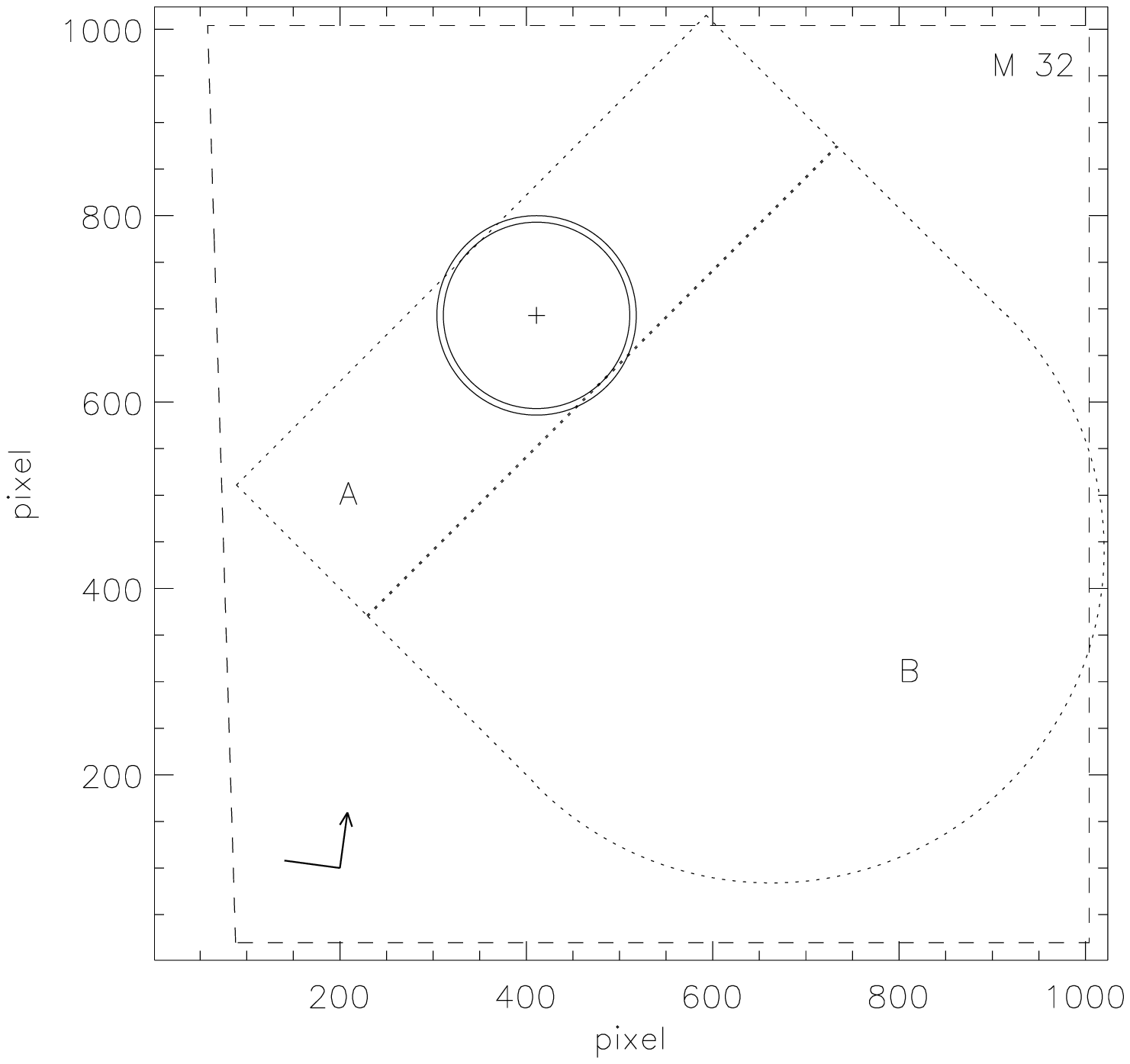}}
\vskip-2.5in
\hspace*{5.0in}
\parbox{2.0in}{
\small {\sc Fig.~\figmbschem--}
The same as Figure~\figmaschem, but for M~32.  Because the M~32 core is
placed closer to the center of our {\it FOC} images 
than was the case with M~31,
we define the {\it IUE} aperture at a different orientation.  
}
\vspace*{0.6in}
\addtocounter{figure}{1}

\clearpage

\parbox{5.75in}{\epsfxsize=5.75in \epsfbox{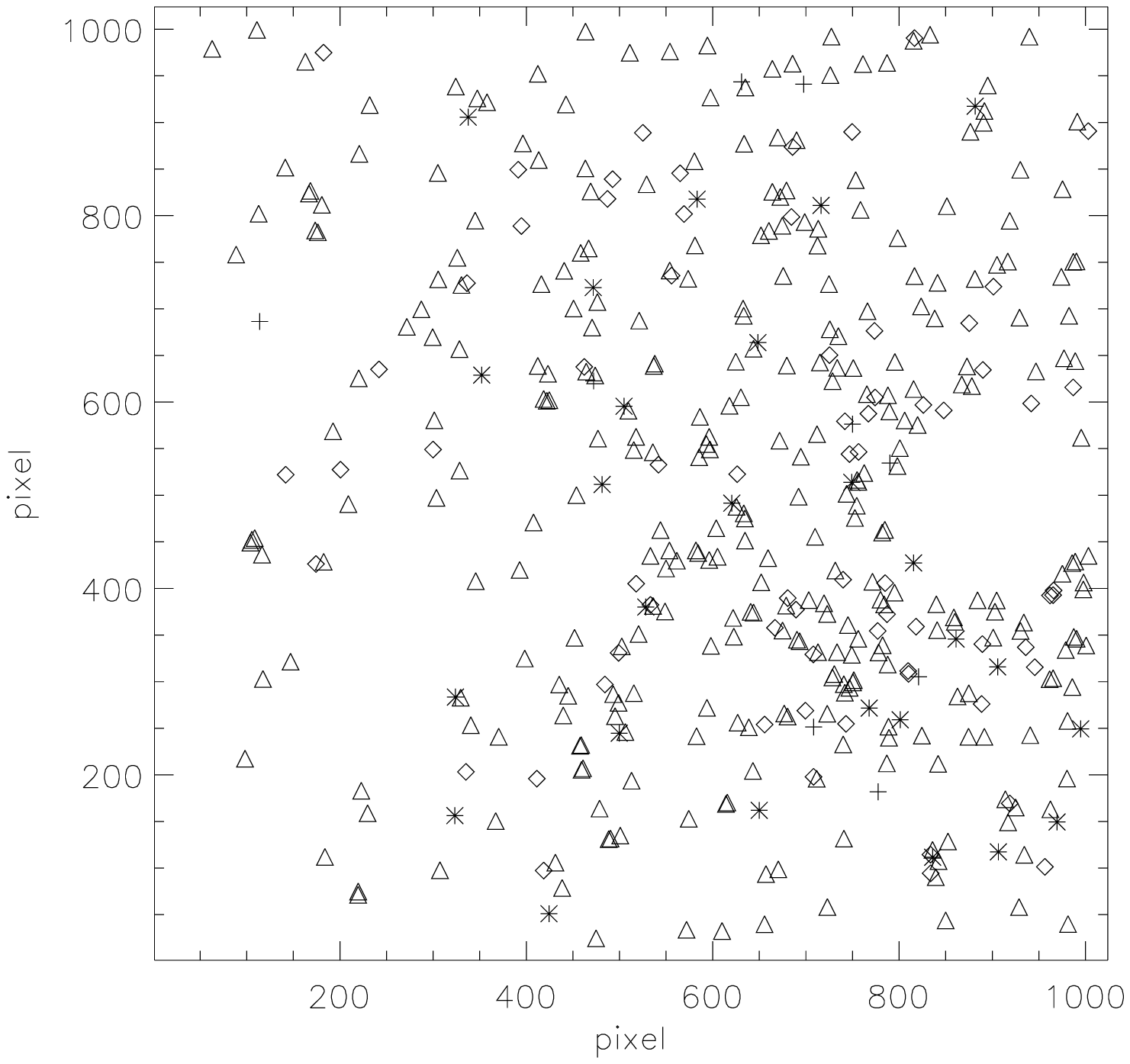}}
\vskip-3.0in
\hspace*{5.0in}
\parbox{2.0in}{
\small {\sc Fig.~\figmamap--}
The photometric catalog of M~31 (Table~\ref{tabm31cat}) plotted on the
{\it FOC} field.  A cross denotes stars with $20.5 \leq m_{F175W} \leq 21.5$,
an asterisk denotes stars with $21.5 \leq m_{F175W}  \leq 22.5$,
a diamond denotes stars with $22.5 \leq m_{F175W}  \leq 23.5$,
and a triangle denotes stars with $23.5 \leq m_{F175W} \leq 24.5$.
}
\vspace*{2.65in}
\addtocounter{figure}{1}

\parbox{5.75in}{\epsfxsize=5.75in \epsfbox{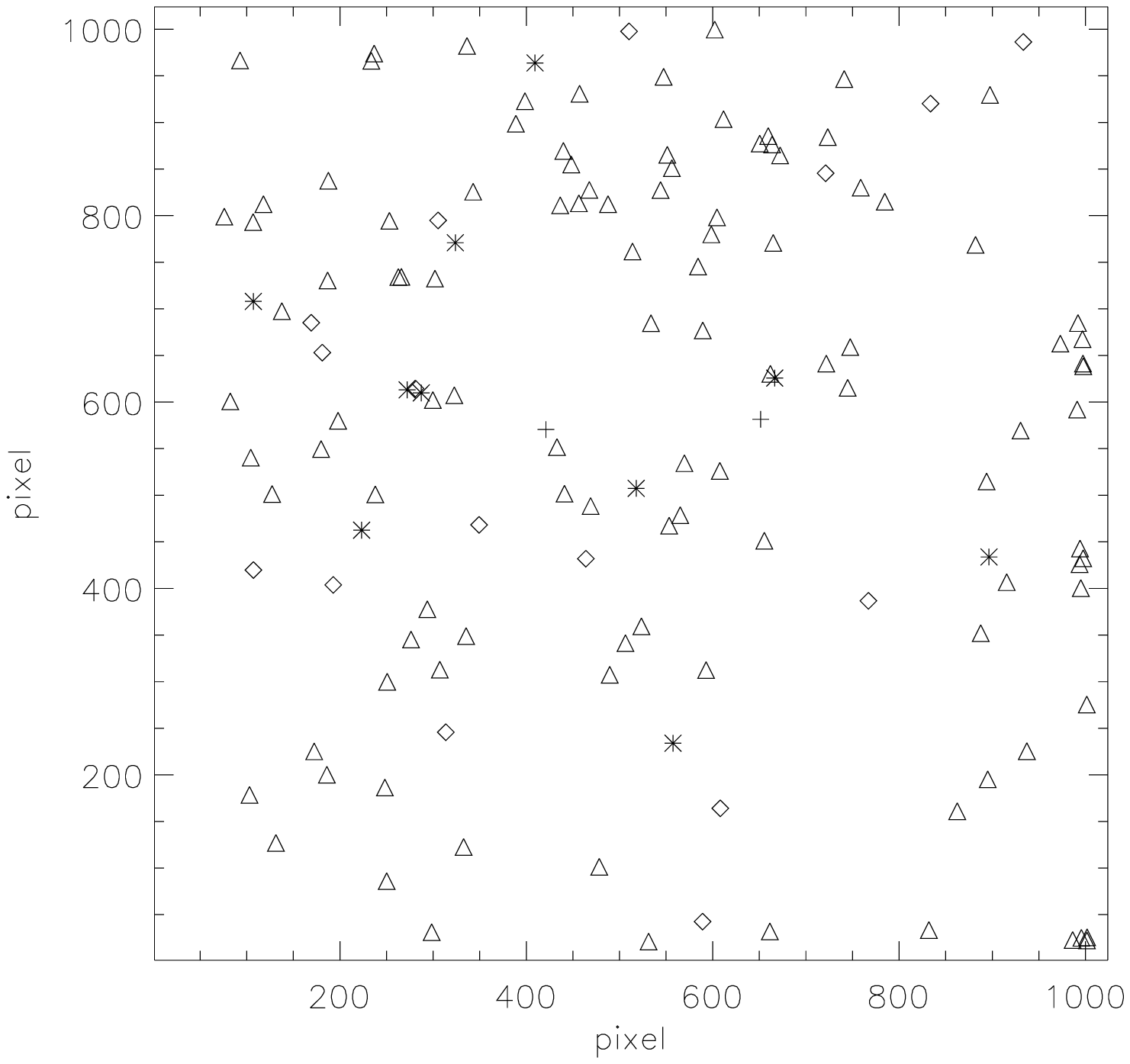}}
\vskip-3.0in
\hspace*{5.0in}
\parbox{2.0in}{
\small {\sc Fig.~\figmbmap--}
The photometric catalog of M~32 (Table~\ref{tabm32cat}) plotted on the
{\it FOC} field, as in Figure~\figmamap.
}
\vspace*{0.6in}
\addtocounter{figure}{1}
\clearpage

\noindent
for each band do not fully correct the
3-pixel flux to the total flux in these stars.  Second, the central
pixel in each star may have ``wrapped'' twice instead of once in two
out of the three F275W frames (they did not wrap at all in the F175W frames).
The additional wraps would require that -0.06 mag be added to the 
$m_{F275W}$ values for these two brightest stars in the table.
The elongation of the stars means that the
aperture corrections for star ``a'' could be -0.97 mag for $m_{F175W}$ and
-0.80 mag for $m_{F275W}$, and the aperture corrections for star ``b'' could be
-0.75 mag for $m_{F175W}$ and -0.68 mag for $m_{F275W}$, instead of the nominal
-0.61 mag and -0.56 mag for $m_{F175W}$ and $m_{F275W}$.  
Because it is unlikely that
these stars are resolved PNe and because the PSF does not seem to vary
with field position (see \S\ref{secobs}), these stars might be merged
images of two or more stars.

We note that the {\it FOC} cameras suffer from nonlinearity at high count
rates (Nota et al.\ 1996\markcite{N96}), and that the M~32 F275W image
may be affected by this nonlinearity in the galaxy core.  For uniform
illumination in this format (F/96 512$z \times 1024$), the maximum count rate
for linearity is 0.04 cts s$^{-1}$ pix$^{-1}$.  Nota et al.\ defines
uniform illumination as varying by ``less than 20\% over scales of 20 pixels.''
The count rate limitation for linearity is much higher in the case of
nonuniform illumination from stars, where the central pixel should not
have a count rate exceeding $\approx$ 1 ct s$^{-1}$ pix$^{-1}$.
The M~32 core is sharply peaked and varies by more than 20\% over
scales of 20 pixels, but the core is not nearly
as peaked as a stellar PSF.  The peak
count rate in the core of M~32 is 0.09 cts s$^{-1}$ pix$^{-1}$, and so
it is likely that there is some loss of counts in the very center
of the M~32 F275W image.  However, given its uncertain nature,
we do not correct for the nonlinearity.
It does not affect our stellar photometry, because we have discarded 
stars within 1.5$\arcsec$ of the galaxy center.

\subsection{Photometric Uncertainties}
\label{secf275}

Our chosen combination of camera parameters
is not frequently used during {\it FOC} observations.
Hence its sensitivity calibration has not been subjected
to as many consistency checks as found in other bandpasses.
Meurer (1995\markcite{M95}) estimates that, in general,
UV photometry with the {\it FOC} is
accurate to $\approx 0.2$~mag.  A review of his Instrument Science Report
(ISR) and other ISRs available on the Space Telescope Web page
(e.g., Jedrzejewski 1996\markcite{J96}) suggests that systematic
errors in calibration of up to $\sim 0.2$ mag are not implausible.

The absolute calibrations of the F175W and F275W filters were done in a
different {\it FOC} format from our observations, and through neutral density
filters (which were calibrated by separate in-flight measurements).
Ground-based and in-flight measurements have shown that the {\it FOC} 
sensitivity depends on the image format, for reasons that are not completely 
understood. The format correction for our observations is a multiplicative
factor of 1.25, as recommended in Greenfield (1994\markcite{G94}).
The major source of uncertainty in the calibration is this format 
correction, which should be wavelength independent.  

Given these uncertainties, we chose to test how well our
{\it FOC} photometry agrees with other measurements. There are several
possible tests:
(1) the magnitudes of the brighter point sources can be compared
to previous {\it FOC} results;
(2) the magnitudes of the brighter point sources can be compared to
photometry of the same sources with {\it WFPC2}; 
(3) the total flux of the M~31 and M~32 nuclei can be compared to 
results from {\it IUE} and {\it WFPC2}.
We discuss these consistency tests in detail below. 

\medskip

\subsubsection{Comparison to Previous FOC results}
\label{secf48}

The current calibration of the {\it FOC} is much more certain
than that of the pre-COSTAR {\it FOC}; the ISRs on the STScI
web page describe the various uncertainties in both the pre-COSTAR
and refurbished {\it FOC}.  The previous {\it FOC} observations of M~31 and
M~32 utilized the pre-COSTAR F/48 camera, and was subject to two large
uncertainties in the calibration.  First, the lack of neutral density
filters in this mode prevented direct calibrations with {\it IUE} standard
stars (as done for the F/96 camera).  Second, the cross-calibrations
between F/96 and F/48 were compromised by the early failure of the F/48
detector.  On top of these uncertainties unique to the F/48 camera, the
photometry in both pre-refurbishment {\it FOC} cameras required large 
encircled energy corrections.

With these uncertainties in mind, 
we compare some of our bright point sources to those found in the M~31 King et 
al.\ (1992\markcite{K92}) analysis.  There are 19 stars
in the King et al.\ photometric catalog (King, private communication)
that can be found in our own F175W image of M~31.
Although our $14 \times 14 \arcsec$ field contains 31 of the stars
found in the $44 \times 44 \arcsec$ field of King et al., the differences
in geometric correction between the old and new cameras
prevent us from securely matching all of the King et al.\ sources to ones
in our own field.
King et al.\ (1992\markcite{K92}) define the correspondence between
their magnitudes and the flux at 1750~\.{A} as:
\[f_{\lambda} = 7.6 \times 10^{-17} \times 10^{-(m-27.60)/2.5}\]
while under the STMAG system, as defined in \S\ref{secphot}, we have 
\[f_{\lambda} = 10^{-(m+21.10)/2.5}.\]

The King et al.\ (1992\markcite{K92}) fluxes are (on average) 
a factor of 5.6 higher than ours for sources in common.  We suspect 
that much of this discrepancy is due to the
difficulty of working with pre-refurbishment data.  Another factor
is the format-dependent sensitivity (Greenfield 1994\markcite{G94}),
which was unknown at the time of the King et al.\ (1992\markcite{K92})
work.  King et al.\ assumed that the FOC sensitivity was 80\% of nominal,
when in reality the zoomed format was 1.44 times {\it more} sensitive.  
Thus, the King et al.\ (1992\markcite{K92}) fluxes
should be multiplied by 0.8/1.44 = 0.56, and the factor of five discrepancy
with our own measurements falls to a factor of three.  In contrast, the
Bertola et al.\ (1995\markcite{BBB95}) recalibration makes these stars
{\it brighter} than determined by King et al.\ (1992\markcite{K92}).
Hence the current {\it FOC} calibrations imply a substantial revision to the
conclusion that the resolved sources in the previous {\it FOC} images
are low-mass PAGB stars.

Although the Bertola et al.\ (1995\markcite{BBB95}) images of M~31 and
M~32 include sources in common with our own images, we find that
their revisions to the pre-COSTAR F130LP+F150W {\it FOC} calibration are 
seriously
in error.  If the sensitivity of the pre-COSTAR {\it FOC} was degraded to the
level claimed by Bertola et al., 61 stars found in their M~31 image would also
be detectable in archival {\it WFPC2} F160BW images of M~31 
(12000 sec exposure) at a S/N ratio of at least 
10 (and as high as 30), assuming the stars to be hot (T$_{\rm eff}$=30000 K)
PAGB stars.  There are no such bright stars in the F160BW image.  In contrast,
if we use the nominal {\it FOC} calibration for the stars detected in
our own F175W image to predict the count rates in the F160BW, we find that
our brightest star would be present at a S/N ratio of 3 in the F160BW, with 
most having a S/N ratio less than 2.  Thus, our photometry is in agreement 
with the F160BW data, which shows no obvious sources
in the field.

Because the previous {\it FOC} photometry is prone to such large errors,
we choose to compare
our data with much more reliable {\it WFPC2} and {\it IUE} measurements, 
as discussed in the next two sections.

\subsubsection{Comparison to WFPC2 photometry of point sources}
\label{secf275wfpc}

M~31 was observed by {\it WFPC2} for 3000 sec through the F300W filter
and 460 sec through the F336W filter.
We reduced the observations following standard procedures, and measured
star magnitudes through a 0.14$\arcsec$ radius aperture,
adopting an aperture correction of -0.24 mag (Holtzman et 
al.\ 1995\markcite{H95}).  There are 48 bright
stars common to all four UV images available (F175W, F275W, F300W, F336W)
in the {\it FOC} and {\it WFPC2} data.  These stars show an average
$m_{F175W}-m_{F275W}$ color of -0.26~mag with an rms dispersion of 0.43 mag.

The {\it FOC} colors correspond to stars with effective temperature 10500~K, 
on the cool end of the populations we are studying
in these {\it FOC} UV images, but in line with expectations when we 
are looking at stars that also appear in the redder {\it WFPC2} filters.
A star with an effective temperature of 10500~K would have a 
$m_{F275W}-m_{F300W}$ color of -0.01~mag, whereas the measured colors average 
0.23~mag (rms = 0.34 mag) for these
stars, i.e., the {\it FOC} magnitudes would need to be 0.24~mag brighter
to produce {\it FOC}-to-{\it WFPC2} colors that agree with the {\it FOC} 
colors.  At the same effective temperature, the 
$m_{F300W}-m_{F336W}$ color should be -0.05~mag, and we measure it to
be 0.06 (rms 0.47~mag).  So, the internal {\it WFPC2} colors of these 48 stars
are more in line with the internal {\it FOC} colors than the 
{\it FOC}-to-{\it WFPC2} colors.
These comparisons would argue that the {\it FOC} photometric zero-point
for our observations is too high.  That is, the F275W magnitudes should
be brighter than we observe by 0.24~mag.  As this is most likely
due to the format dependence, such a correction would affect both {\it FOC}
bands.  We shall see below that such a correction to the F175W sensitivity
would put the F175W fluxes of the two galaxy nuclei rather seriously 
in disagreement with the {\it IUE} results.

Taken at face value, the red $m_{F275W}-m_{F300W}$ colors would
imply stars of effective temperature 5000 K to 10000 K. 
This is inconsistent with the blue {\it FOC} colors for the same sources and
with their non-detection in the {\it WFPC2}.
For a star of 5000~K, the count rate in F814W would be 75 times
higher than in F300W, and all of these stars would be detected at
signal-to-noise ratios of 5--170 in the F814W image. 
These stars are no brighter than the 5 $\sigma$ level at their positions in 
the {\it WFPC2} F814W image.

\subsubsection{Comparison to IUE and WFPC2 Photometry of the Nuclei}

The integrated fluxes of the central regions of the two galaxies in
our {\it FOC} images can be compared to photometry from {\it WFPC2} and to
fluxes measured by {\it IUE}.  The comparison is complicated by the 
large corrections that need to be applied to compare the {\it IUE}
and {\it FOC} fluxes in equivalent aperture sizes.

Unsaturated archival {\it WFPC2} data for the M~31 and M~32 cores exist
in the F160BW, F300W, F336W, F555W, and F814W bands.  Unfortunately, in the 
F160BW filter, M~32 is invisible and M~31 is barely detectable, so we 
discarded these far-UV {\it WFPC2} data and used the other {\it WFPC2} bands.
Composite UV+Optical spectra of the cores of M~31 (McQuade, Calzetti,
\& Kinney 1995\markcite{MCK95}) and M~32 (Calzetti, private communication)
were fed through the IRAF package SYNPHOT to determine the expected colors 
of the nucleus as seen in the {\it FOC} and {\it WFPC2} bands
available to us (Table~\ref{tabf275}).  
These composite spectra were spliced together from
aperture-matched {\it IUE} and ground-based data.  The {\it IUE}
aperture sampled a large $10 \times 20\arcsec$ area centered on each
galaxy nucleus.  We extended these spectra beyond 7500~\.{A}, to
12000~\.{A}, by defining a flat continuum that reproduced the {\it WFPC2}
$m_{F555W}-m_{F814W}$ colors in each galaxy.  This red extension for
each spectrum was only needed to reproduce the red leak
contributions properly for the various {\it FOC} and {\it WFPC2} filters.

After using SYNPHOT to compute the predicted {\it FOC} and {\it WFPC2}
count rates for a $10 \times 20\arcsec$ aperture centered on the
nucleus, we measured the count rate in the archival {\it WFPC2} data
and our own {\it FOC} data, using an aperture with a diameter of
2.8$\arcsec$ centered on the galaxy cores (see
Figures~\figmaschem\ and \figmbschem).  
We chose an aperture
of this smaller size to reduce the systematic errors introduced from
our background assumptions.  
The SYNPHOT predicted count rates were then normalized to agree in the
visual (F555W), accounting for the difference in aperture sizes.
In Table~\ref{tabf275} we show the net count rate in this aperture for
each of the images, as well as the gross count rate, dark
count rate, and sky count rate.  Dark counts
were subtracted from the {\it WFPC2} data in the pipeline processing
(hence the zero values in the table); for the {\it FOC} dark counts, we
used the nominal value of $7\times 10^{-4} \rm~cts~s^{-1}~pix^{-1}$
from Nota et al.\ (1996\markcite{N96}), and this agrees well with the
value determined independently from those portions of the detector
occulted by the {\it FOC} ``fingers'' (see Figures~\figma\ \&
\figmb).  The sky contributions came from running SYNPHOT with
the STScI sky template for high zodiacal and high Earth shine.  We
chose high sky values because of the small sun-angle during these
observations ($\approx 60^{\rm o}$);  however, because of the small
aperture we are using (2.8$\arcsec$ diameter), choosing lower values of
sky background would produce negligible changes in our results
(Table~\ref{tabf275}).  We note that the {\it WFPC2} data were cosmic
ray rejected using the IRAF package CRREJ, and were also corrected for
hot pixels using the IRAF package WARMPIX.

In Table~\ref{tabf275}, comparison of the 
predicted {\it IUE} count rates in each {\it HST} band vs.\ the net measured
count rates suggests a discrepancy in the F275W flux, in the
same sense as that discussed above. That is, it appears that the flux
in F275W is about 25\% lower than expected, based on the {\it WFPC2}
observations.  However, it appears that the F175W flux is reasonably
close to expectations based upon {\it IUE} and {\it WFPC2} measurements.
There is a radial color gradient in each of these galaxies, so we do not 
expect a completely flat net/predicted (2.8$\arcsec$/10$\times$20$\arcsec$)
count rate ratio all the way from the UV to the visual; it is evident
in Table~\ref{tabf275} that these galaxies become redder with increasing
radius.  However, the F275W count rate is noticeably shifted with respect 
to the neighboring F175W count rate (in the {\it FOC}) and the F300W count 
rate (in the {\it WFPC2}).  The M~31 F275W net/{\it IUE} count rate ratio 
requires multiplication by 1.25 in order to bring it into agreement with the 
ratio in the neighboring bands,
and the M~32 F275W net/{\it IUE} count rate ratio requires multiplication
by 1.43.  However, the central region of the M~32 F275W 
image has a count rate that probably falls in the nonlinear regime for 
{\it FOC} extended sources (see \S\ref{secphot}), and this effect
may be decreasing the number of counts measured in the 2.8$\arcsec$ aperture; 
the M~31 nucleus should not be suffering from this nonlinearity.  
Hence, the data in Table~\ref{tabf275} might
suggest that the nominal F275W values of PHOTFLAM 
(9.38$\times 10^{-18}$) and $m_o$ (21.47 mag) should be respectively
multiplied by 1.25 and shifted by -0.24 mag in order to agree
with predictions from {\it IUE}.  

To summarize, we have compared our {\it FOC} photometry to photometry
from {\it WFPC2} archival images and to spectrophotometry from {\it IUE}. We
find evidence for systematic discrepancies at the 0.25~mag level.
However, applying such a correction
would produce stellar colors that are incompatible with {\it WFPC2} 
photometry and with plausible stellar evolutionary tracks.  As we cannot
identify an obvious flaw in the sensitivity calibration 
(Jedrzejewski 1996\markcite{J96})
that would lead to a systematic difference between F275W and F175W, we
have chosen to adopt the pipeline {\it FOC} sensitivity calibration with
no additional corrections.  We must caution that our interpretation rests
rather heavily on this calibration.  In a sense, we have taken the
approach of adopting the nominal calibration to see where it leads, in 
spite of the disturbing inconsistencies with other observations.
Further calibrations of the {\it FOC} and/or observations of these galaxies
with {\it STIS} will be required to sort out the calibration issues
with greater certainty.

\subsection{Completeness and Spurious Sources}
\label{seccompspur}

At the 4~$\sigma$ detection limit, the faint sources in our catalog 
represent only a small fraction of those actually present in the population.
Furthermore, many spurious sources are detected through local maxima in the 
noise of the background.  To properly characterize the completeness of our
photometric sample and the contamination from spurious sources, we
ran photometry simulations of the {\it FOC} data.

To characterize the completeness for a given band in one of the galaxies, 
we created a blank image frame and added 10 stars (one for each magnitude from
21 to 30) at random positions using the PSF sizes listed 
in Table~\ref{tabobs}.
We then block summed the image along one axis (combining every 2 pixels
into one), added Poisson noise, dezoomed the image back to $1024 \times 1024$
pixels, and applied the geometric correction for the {\it FOC}.  This process
gave the ``fake'' stars the proper distortion and noise characteristics.
We then added this image of fake stars to the true {\it FOC} image of the 
galaxy, and performed our source detection and photometry to attempt recovery
of these fake stars.  The entire simulation
was repeated 100 times, and the resulting photometric records allowed
us to determine the completeness of our sample as a function of the
photometric magnitude.  These results are shown in Table~\ref{tabcompspur},
along with the standard deviation in the recovered magnitudes.
The recovered magnitudes for the fake stars also 
provided an independent measurement
of the aperture correction, and this was in agreement with the correction
determined from direct measurement on isolated stars (\S\ref{secphot}).

The spurious source characterization proceeded in a similar manner.
Starting with the actual galaxy image, we corrected the flux between the 
edge of the detector and the image border to match that on the detector, 
and then applied a $35 \times 35$ median filter to
remove all of the stars from the image.  We then block summed this image
along one axis, added Poisson noise, dezoomed, and applied the 
geometric correction.  This last step of applying the geometric correction
restores the gap between the detector edge and the image border.
Next, we applied our source detection and photometry
to this image and recorded the number of stars found at each magnitude from
21 to 30.  The simulation was repeated 100 times, and allowed
us to determine the average number of spurious sources found at each
magnitude (the median filtering of the image meant that any detected source,
by definition, would be spurious).  
We took the average number of spurious sources and divided by the actual
number of sources found in that galaxy image to determine the
contamination from spurious sources (Table~\ref{tabcompspur}).
The contamination at very faint magnitudes exceeds 100\% because of 
slight differences between the simulation ``sourceless'' images and the
real image.  For example, a fraction of the area in the real image is
taken up by the PSFs of bright stars that are not present in the sourceless
images.  Also, the median filter will slightly redistribute the flux
in the bright stars to the background.  Finally, the background in the
galaxies is comprised mainly of horizontal branch stars that are
below the detection limits, presenting a non-uniform background.
Thus, although the simulation
sourceless images have very nearly the same background characteristics
as the real images, they are not exactly the same, and so the
spurious source contamination at faint magnitudes only serves as a guide
for the limits to our real photometry.

As evident from Table~\ref{tabcompspur}, 
the completeness and spurious source contamination
is very similar for the two galaxies, while a comparison
of the different bands shows that the F275W band reaches fainter than
the F175W band.  Based upon our simulations, we find that our photometry
is reliable down to $m_{F175W}$=24.5 mag and 
$m_{F275W}$=25.5 mag, beyond which the number of detected sources
drops rapidly and the contamination from spurious detections
increases significantly. 

Because the galaxy background varies dramatically over the entire image,
and is very large in the cores of the galaxies, our completeness and
spurious source contamination is a function of position
in each image, and the spurious sources in the very centers of the
galaxies are very bright.  These bright spurious sources
will not be rejected by discarding stars fainter than $m_{F175W}$=24.5 mag and 
$m_{F275W}$=25.5 mag; the limits only represent the average characteristics
of the image.  Color magnitude diagrams that include the 
inner 1.5$\arcsec$ of each galaxy thus show a well-resolved group of bright 
sources, separate from the rest of the population.
We assume that these sources are spurious (and not an 
unusual class of stars residing in the galaxy cores), and so 
we discard sources that lie within
1.5$\arcsec$ of the center of each galaxy.  Our completeness and spurious
source contamination, shown in Table~\ref{tabcompspur}, 
reflects the discarding of these sources.  

It is important to note that our completeness and spurious source
contamination has been defined using detection and photometry in
a single band.  Without making assumptions about the colors of the
stars we are simulating, this is the only way to characterize our 
photometry.  However, our photometric catalog includes more faint stars in each
band than expected from the completeness simulations, because a star that
is fainter than a 4~$\sigma$ detection in one band might have been 
detected in the other if it was bright
in the other band.  For this reason, our luminosity functions (LFs) in each
band only include sources that were detected in that band, as discussed
in \S\ref{seclf}.  

\section{ANALYSIS}

\subsection{Distribution of Diffuse and Stellar Light}

If the UV-bright stars are drawn from a population that is different from
the diffuse population (either in age or metallicity), differences
might be detected in the distribution of diffuse light and stellar light 
in these galaxies.  In Figure~\figdist, we show the distribution of 
diffuse light as compared to the distribution of light from bright stars.  

We define bright stars as those detected at luminosities 1~mag brighter than 
the detection limits in a given band;
e.g., the F275W bright stars are those stars detected in F275W with
magnitudes brighter than $m_{F275W} = 24.5$~mag.
Fainter stars were excluded from the analysis to minimize contamination
from spurious sources and to ensure that the completeness of the sample
did not vary with position in the images.
To measure the diffuse light in the {\it FOC} images,
we removed a 4-pixel radius area around each of the bright stars, 
replacing it with the median value in the area outside
of this radius and bounded by a $15\times 15$ pixel square.
The diffuse light in each annulus was then measured after subtracting the
dark counts and sky background (assuming high zodiacal and high Earth shine,
as discussed in \S\ref{secf275}).
The distributions of diffuse light and bright stellar light 
were measured in annuli chosen
to keep the number of bright stars in each annulus constant; in this way, the
statistical uncertainty in each annulus (due to the small number of bright
stars) can be kept approximately constant.
There are 25 bright stars in each M~31 annulus, and 7 in each M~32 annulus.

Although the distributions of stars are noisy, it is
apparent that the light from UV-bright stars is more tightly 
concentrated toward the galaxy centers than the diffuse light.
The trend has also been seen in previous studies (cf.\ Bertola et 
al.\ 1995\markcite{BBB95}).
It is well-known that metallicity gradients exist in early-type galaxies,
in the sense that the metallicity decreases with distance from the center
(cf.\ Gonz$\acute{\rm a}$lez \& Gorgas 1995\markcite{G95} and references 
therein).  If the UV-bright post-HB stars are more easily produced at
higher metallicities, it would make sense that they would fall off
more rapidly with radius than the underlying diffuse light.  This is 
because the diffuse light has contributions from the main sequence turnoff
and any metal-poor HB population (if present).  The luminosity
of the main sequence is affected by the metallicity to a much smaller
degree, and in the opposite sense (i.e., it becomes brighter at decreasing
metallicity); the metal-poor blue HB population would become more prominent 
as we move outward with decreasing metallicity.

\medskip

\parbox{3.0in}{\epsfxsize=3.0in \epsfbox{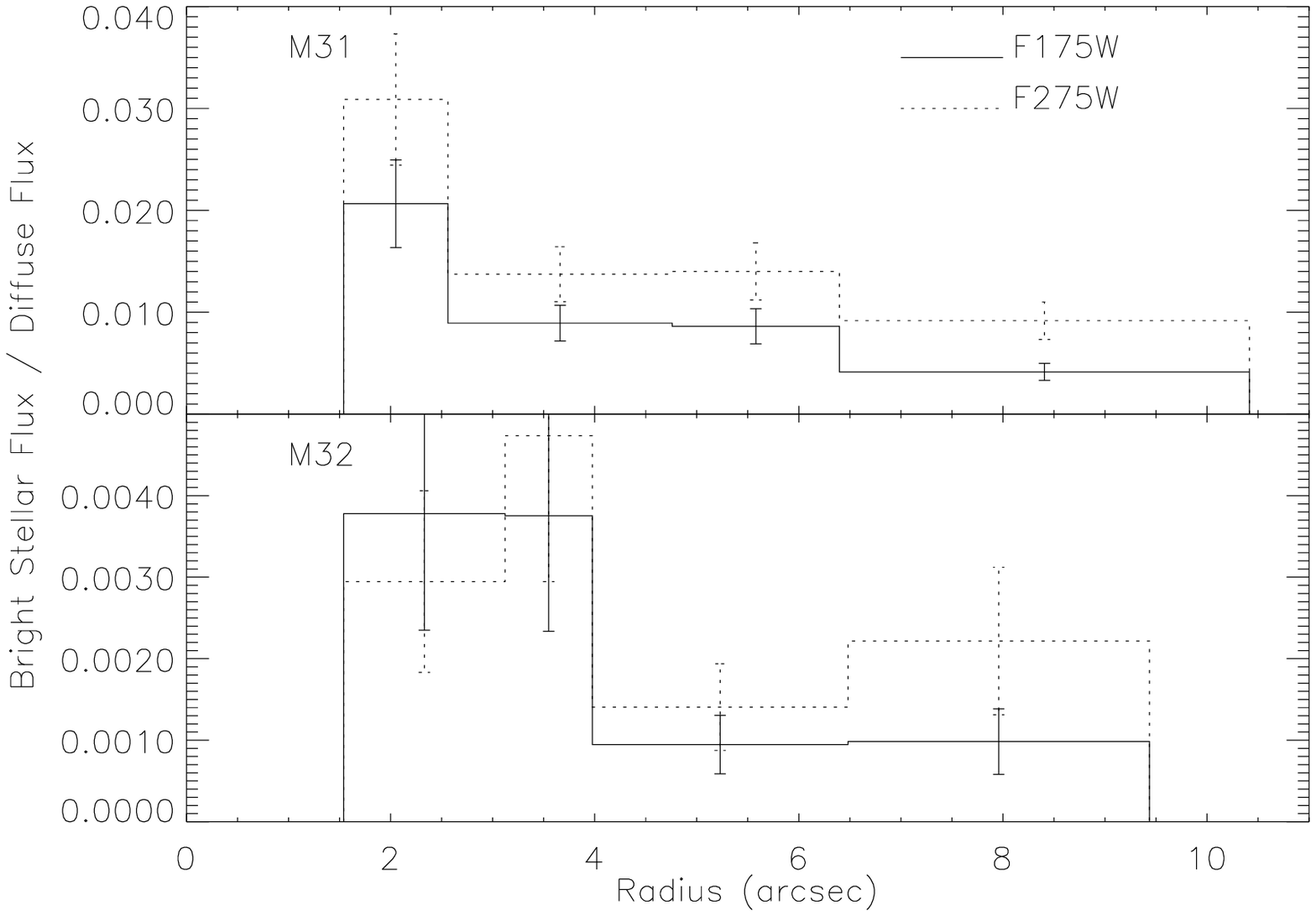}}
\vskip 0.1in
\centerline{\parbox{3.0in}{\small {\sc Fig.~\figdist--}
The ratio of flux in bright stars to the flux in diffuse light shows
a rapid decline with radius in the galaxies.  UV-bright stars are defined
as those 1~mag brighter than the detection limits in each band, and were
chosen to ensure that the completeness remained constant with position,
and also to reject any spurious sources from our analysis.  The sizes
of the annuli were chosen to keep the number of stars constant in 
each annulus, in order to provide fairly uniform statistical uncertainties
from bin to bin.  Each M~31 annulus contains 25 UV-bright stars, and each
M~32 annulus contains 7 UV-bright stars.  Error bars reflect statistical
uncertainty in both the number of stars and counts, but are dominated
by the number of stars.
}}\vspace{0.5in}
\addtocounter{figure}{1}

\subsection{Luminosity Functions}
\label{seclf}

The photometric analysis on each galaxy yielded catalogs of 
stars in M~31 and M~32.  However, as noted in 
\S\ref{secdet}, the catalog in each galaxy
is made up of many stars that were detected (at 4~$\sigma$) in only one band 
but have photometry in both bands.  Without making assumptions regarding
the color of the population, we can only characterize our detection limits 
using simulations in one band (see \S\ref{seccompspur}), so we create a 
luminosity function for a given band by drawing upon only those stars in our 
catalog that were detected in that band.  Knowledge of the star from the 
other band is ignored in this analysis, and so, for example,
a bright star in F175W that is fainter than $m_{F275W}$~=~25.5 mag will 
be included in the F175W histogram.  In that way, we can compare histograms
while using what we know about the completeness and spurious source
contamination.

The raw LF for each galaxy in each band is shown in Figure~\figlf.
Although we could correct for the completeness 
and spurious source contamination (Table~\ref{tabcompspur}) to plot
a ``corrected'' LF for each band, this is not
necessary, because the completeness and spurious source contamination are
approximately the same for the different galaxies in each band.
The raw LFs can be directly compared without these corrections.

\begin{figure*}[t]
\parbox{7.0in}{\epsfxsize=7.0in \epsfbox{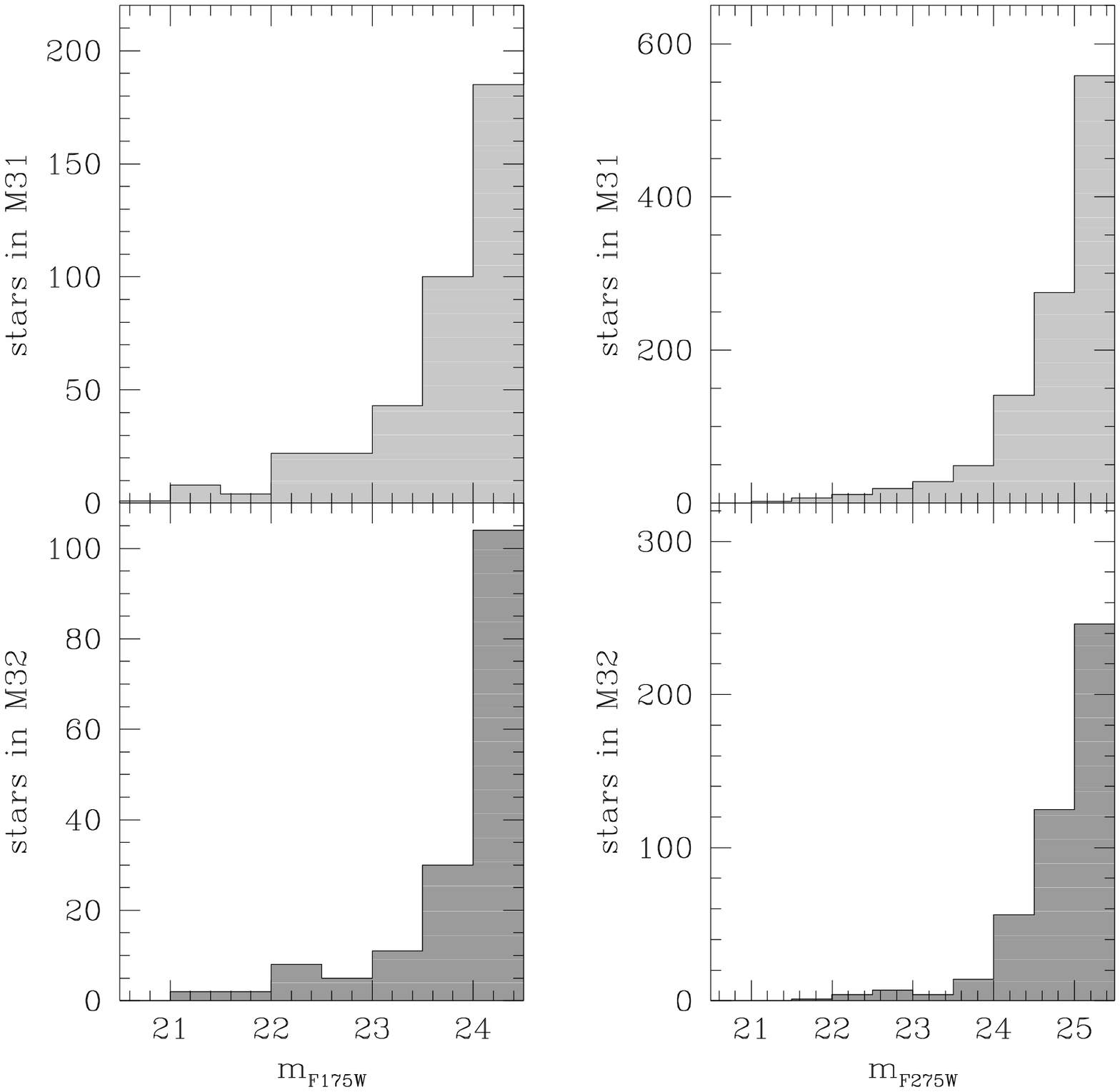}}
\caption{
The raw luminosity functions for each image shows that the LFs for M~31 
(upper panels) look similar to the LFs for M~32 (lower panels).  
These luminosity functions
have not been corrected for completeness and spurious source contamination,
because these characteristics are the same for each galaxy in a given band.
The luminosity function for a given band only draws upon those sources in
the catalog that were detected in that band.  The LF for a given band
also includes those stars
not found in our catalogs that were fainter than the detection limit in
the other band.  Note that the scale in
the M~32 luminosity functions is half that in the M~31 luminosity functions.
}
\end{figure*}

\begin{figure*}[t]
\centerline{\parbox{5.0in}{\epsfxsize=5.0in \epsfbox{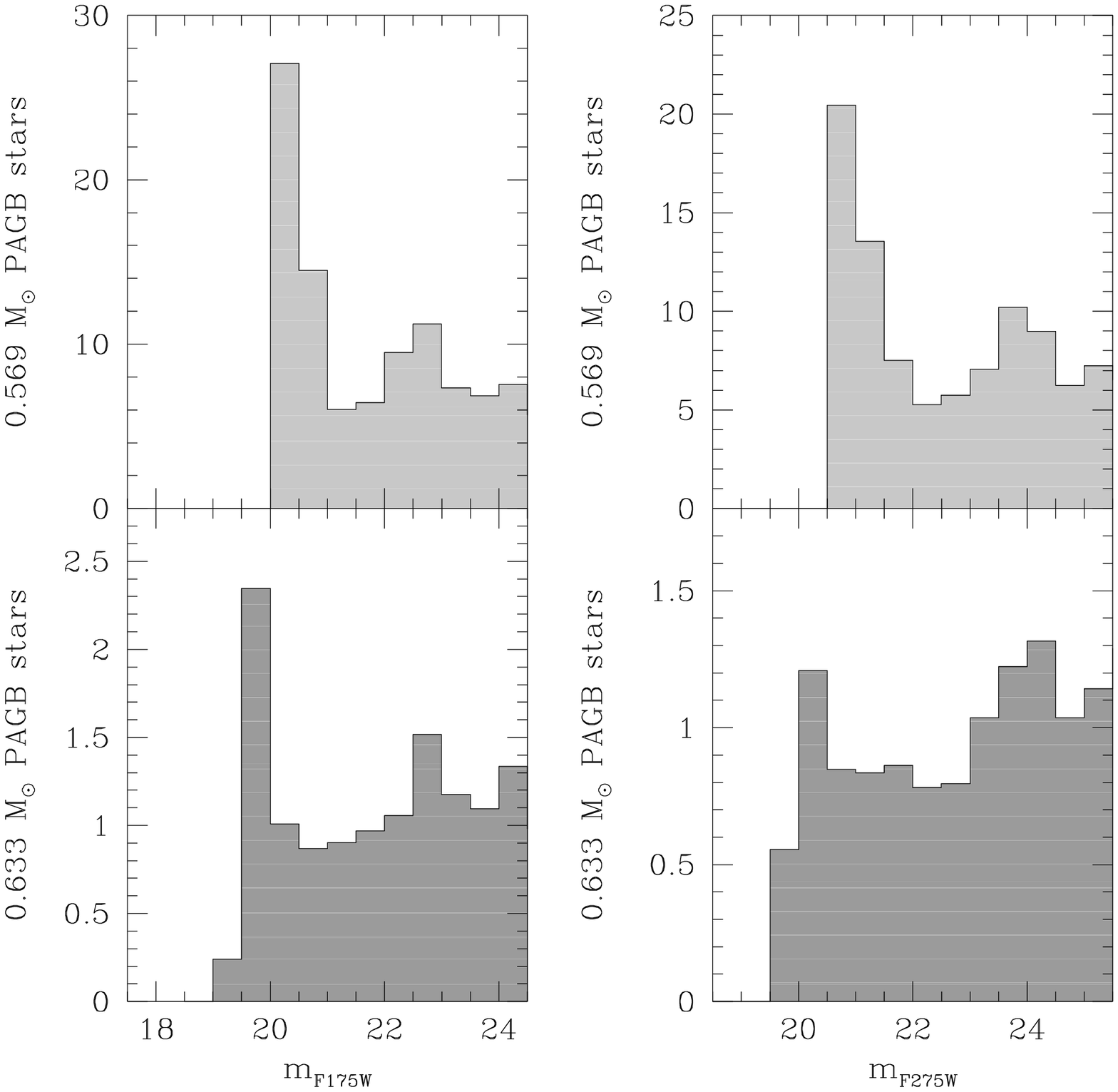}}}
\caption{
The luminosity functions for stars entering a 0.569~M$_{\odot}$ PAGB
track (top panels) and a 0.633~M$_{\odot}$ PAGB track (bottom panels)
show that PAGB stars are not the dominant component to the
populations detected in our {\it FOC} images.
The scaling was chosen to match
the population size
allowed under the fuel consumption theorem in the center of
M~31; the 0.633~M$_{\odot}$ stars evolve more rapidly and so fewer would be 
found above our detection limits at a given evolutionary rate.
The characteristic shape of these
LFs (flatter than we observed, with peaks significantly above our detection 
limits) would be easily detected in the M~31 and M~32 data, implying that
our detected UV sources are not dominated by PAGB stars.
}
\end{figure*}

The LFs in a given band look similar when we
compare the two galaxies.  None of the LFs peak before the sample
becomes seriously incomplete and contaminated, but a comparison of the
slopes shows no dramatic differences between the stellar populations
in each galaxy.  From the LFs alone, we could conclude that the 
UV-to-optical light ratio is weaker in M~32 compared to M~31 simply
because there are about half as many UV sources in M~32, and not because 
these sources are derived from a different population.

Our LFs demonstrate that the UV-bright populations in M~31 and M~32 are not
dominated by PAGB stars (assuming that they are free of strong circumstellar
extinction).  If all of the UV-bright stars were following PAGB 
tracks, we would expect to observe bright peaks in the luminosity function, 
well above our detection limits; instead, we see LFs that rise sharply and
monotonically at decreasing luminosity.  For example, we show in 
Figure~\figpagblf\ synthetic luminosity functions for a population of
stars entering a low-mass (0.569 M$_{\odot}$) PAGB track and 
an intermediate-mass (0.633 M$_{\odot}$) PAGB track (Vassiliadis \& Wood 
1994\markcite{VW94}) at the maximum rate allowed in the
center of M~31 under the fuel consumption
theorem.  For old stellar populations (age~$\geq 10^{10}$~yr), 
the number of stars evolving through any evolutionary phase cannot exceed
2.2$\times$10$^{-11}$ stars L$_{\sun}^{-1}$ yr$^{-1}$ 
(Greggio \& Renzini 1990\markcite{GR90}).  We compute the bolometric 
luminosity in a given aperture by measuring the F555W {\it WFPC2} flux, 
correcting for the 0.06~mag color offset between the F555W filter and the 
Johnson $V$ band, and using a bolometric 
correction (BC) of $-1.25$.  This BC is consistent with a stellar
population of age 12~Gyr and [Fe/H]~=~0.25 (Worthey 1994\markcite{W94}),
and was chosen on the assumption that the centers of M~31 and M~32 are
dominated by old, metal-rich populations.  Younger populations would
have a smaller BC (e.g., $-1.13$ for an age of 8~Gyr), as would populations 
with less metallicity (e.g., $-0.96$ for [Fe/H]~=~0.0), but these effects are 
not very significant for our purposes here. 
The magnitudes in each band are computed with the IRAF package SYNPHOT, using
appropriately reddened and normalized Kurucz (1992\markcite{KU92}) synthetic 
spectra.  The PAGB tracks begin after the thermally pulsing (TP) phase, 
at T$_{\rm eff}$~=~10000~K, and so miss the UV-faint contribution from the 
rapid evolution up the AGB and through the TP, but qualitatively the LFs should
be valid.  Circumstellar reddening in the early phases of PAGB 
evolution could alter the appearance of a PAGB-dominated LF if the
stars are hidden behind dust shells during their earliest and brightest
phases (see \S\ref{secext}).   Figure~\figpagblf\ also demonstrates 
that only a dozen or so intermediate-mass PAGB stars could be detected
in our M~31 images, while approximately 100 longer-lived low-mass PAGB stars 
could be detected in these images.  The stellar population in M~31 and
M~32 cannot be dominated by low-mass PAGB stars, because they would be
detected in significant numbers in our images, unless these stars
are hidden by circumstellar extinction (see \S\ref{secext}).  More massive 
PAGB stars ($\geq$ 0.633 M$_{\odot}$) could dominate the stellar population in
these galaxies but not that fraction of the population {\it above our detection
limits,} because they rapidly evolve through their brightest phases.
We stress this distinction between PAGB domination of the detected
UV-bright population, and PAGB domination of the {\it entire} 
population, which is certainly plausible (and indeed likely, as explained
in \S\ref{seciue}).  A survey using a wider field of view, perhaps with
{\it STIS}, would be able to detect intermediate-mass 
(0.633 M$_{\odot}$) PAGB stars if the
survey reached as deep as our {\it FOC} data; for example, given a program
that sampled a $100 \times 100 \arcsec$ field, 296 PAGB stars would be detected
in M~31 and 52 would be detected in M~32.

\subsection{Color Magnitude Diagrams}
\label{seccmd}

\begin{figure*}[t]
\parbox{7.0in}{\epsfxsize=7.0in \epsfbox{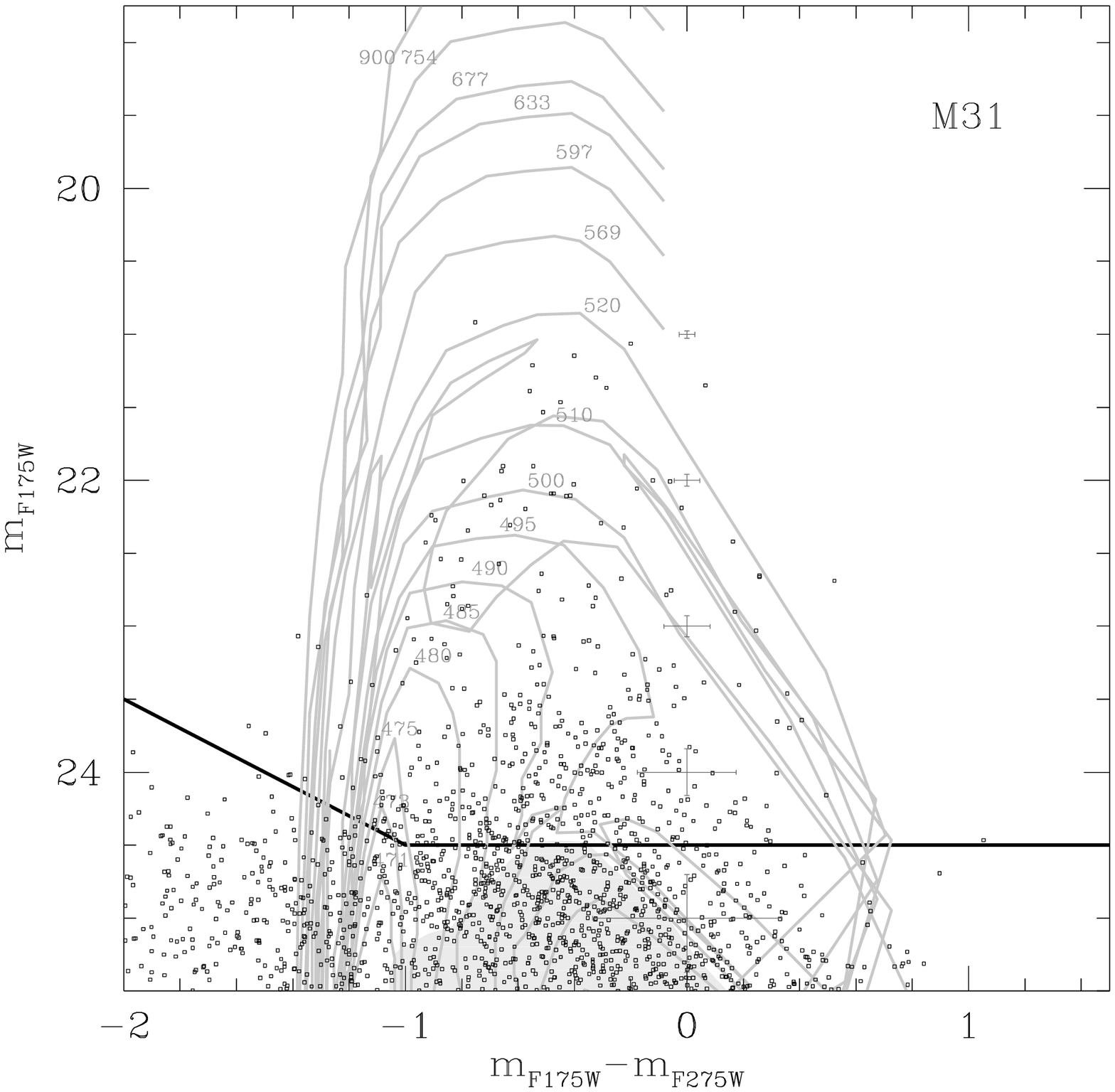}}
\caption{
The color-magnitude diagram for M~31 shows stars (boxes) arising from a broad
distribution of masses (labels) on the HB.  The post-EHB evolutionary paths
from Dorman et al.\ (1993\protect\markcite{DRO93}) and the
PAGB paths of Vassiliadis \& Wood 
(1994\protect\markcite{VW94}) are plotted in light grey.  Stars should avoid
the shaded region where the evolution is relatively rapid, but
this region falls below our limiting magnitudes (solid black line) and
is also blurred out by the large statistical uncertainty (error bars)
for these faint magnitudes.  The AGB is not shown
due to uncertainties in the mass loss on the AGB
and in the thermally pulsing stage at the end of the AGB.
}
\end{figure*}

\begin{figure*}[t]
\parbox{7.0in}{\epsfxsize=7.0in \epsfbox{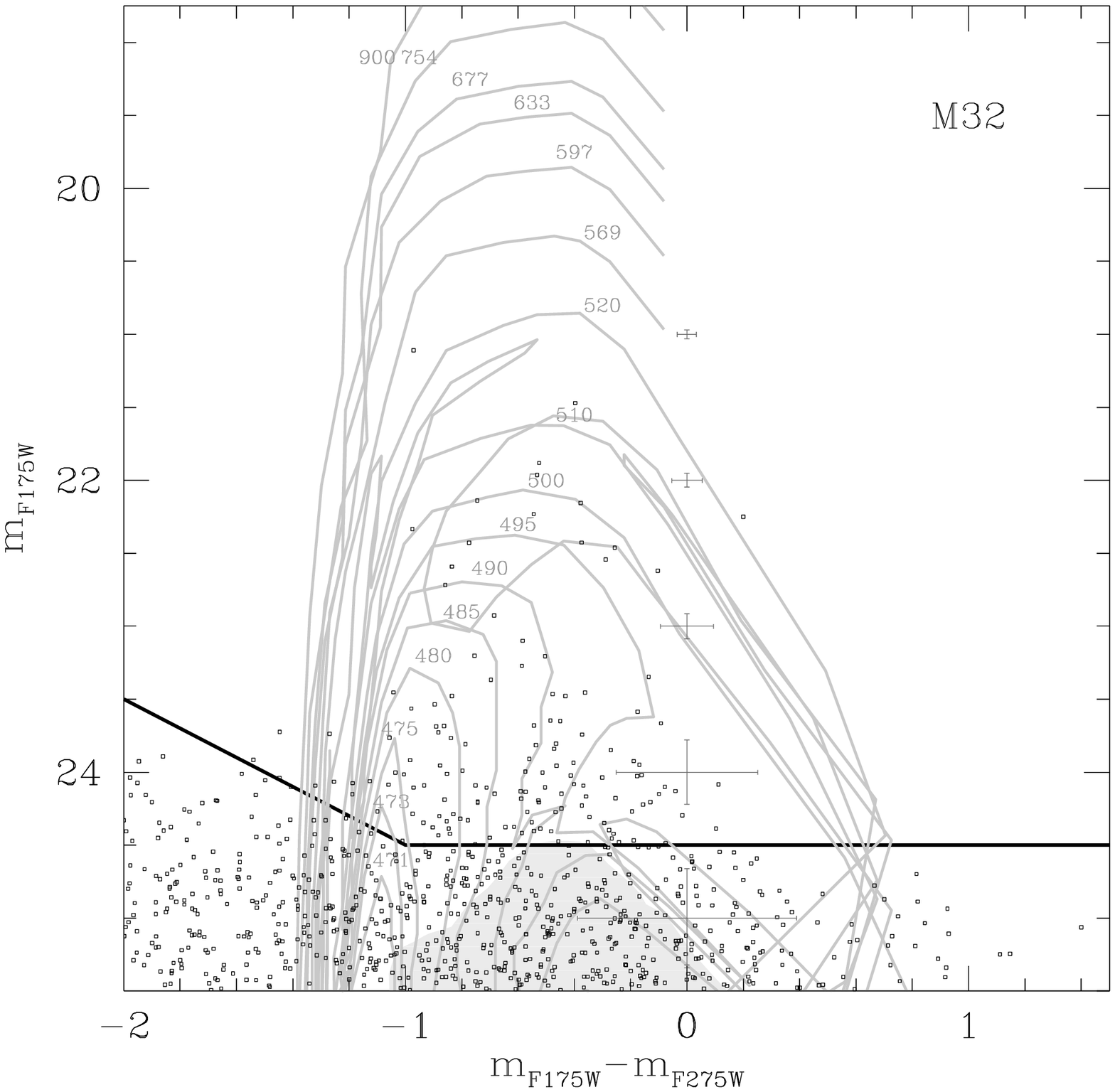}}
\caption{
The color-magnitude diagram for M~32 appears very similar to the M~31
CMD (Figure~\figmacmd).  The only notable distinctions here are the
fewer number of stars and the lack of red stars.
}
\end{figure*}

We plot the color-magnitude diagrams for M~31 and M~32 in 
Figures~\figmacmd\ and \figmbcmd.  In each figure, we also
plot the evolutionary tracks of Dorman, Rood, \& O'Connell,
(1993\markcite{DRO93}) 
for solar-metallicity post-EHB stars (0.471--0.520 M$_{\odot}$) and 
the evolutionary tracks of Vassiliadis \& Wood (1994\markcite{VW94})
for solar-metallicity H-burning PAGB tracks (0.569--0.900 M$_{\odot}$).  
The AGB evolution prior to the Vassiliadis \& Wood tracks is not shown,
due to uncertainties in the thermally pulsing phase and
mass loss on the AGB.  Although
we plot all sources down to $m_{F175W} = 25.5$~mag, the thick black line
denotes the limiting magnitudes as determined
in \S\ref{seccompspur}.  The errors for a star of a given magnitude
vary depending upon the local sky background, and so we show average
1~$\sigma$ error bars as a function of photometric magnitude.
The hottest EHB track (0.471~M$_{\odot}$) plotted has an envelope mass of
0.002~M$_{\odot}$, and at its brightest is too faint to be detected
in our {\it FOC} images.  Stars of lower envelope mass are possible
and are detected in the Galactic field (cf.\ Mitchell
et al.\ 1998\markcite{M98} and references therein), 
but were not included in the models of Dorman et al.\ (1993\markcite{DRO93})
due to computational limitations.

The {\it FOC} magnitudes in the tracks are computed by taking the closest 
Kurucz (1992\markcite{KU92}) synthetic spectrum for the given evolutionary 
track step, correcting the absolute flux level to account for the mismatch in 
temperature and gravity between the Kurucz grid and the evolutionary track,
applying reddening (Cardelli, Clayton, \& Mathis 1989\markcite{CCM89})
due to extinction in our own Galaxy (see Table~\ref{tabobs}), 
normalizing as appropriate to the distance (see Table~\ref{tabobs}), 
and then running the IRAF STSDAS package SYNPHOT.  
We note that the extinction is a large source of uncertainty in our
analysis.  For the foreground extinction due to reddening in our own Galaxy,
values in the literature vary from $0.035 \le E(B-V) \le 0.11$ mag
(Tully 1988\markcite{T88}; McClure \& Racine 1969\markcite{MR69}; Burstein et 
al.\ 1988\markcite{B88}; Ferguson \& Davidsen 1993\markcite{FD93};
Burstein \& Heiles 1984\markcite{BH84}).  
Furthermore, internal extinction in the centers of M~31 and
M~32 is certainly plausible (cf.\ Bianchi et al.\ 1996\markcite{BCBHM96}
and Ciardullo et al.\ 1988\markcite{C88}),
and may even be patchy in the center of M~31.
A fraction of these stars may also have circumstellar extinction
from dust shells ejected during their evolution (e.g., as planetary nebulae);
the timescale for the ``thinning'' of such circumstellar extinction 
(at least in the optical bands) is very uncertain.
We explore PAGB circumstellar extinction in \S\ref{secext}.

The most important aspect to note about this diagram is that the
tracks leaving the blue extreme of the horizontal branch
appear to be populated.  These post-EHB stars are the same hot subdwarfs
seen in our own Galactic field, and our {\it FOC} images represent the
first time these stars have been directly detected outside of our own
Galaxy. 

As stars evolve from the HB and toward eventual death on the white
dwarf cooling track, they spend the longest fraction of time on the
HB itself (note that the HB falls below the bottom of Figures~\figmacmd\ \& 
\figmbcmd).  Upon core helium exhaustion, stars will either
ascend the AGB (not shown in the figures) or one of the post-EHB
tracks (following either AGB-Manqu$\acute{\rm e}$ or post-early-AGB
behavior).  The evolution upon leaving the HB is relatively
rapid, but then slows down again, before proceeding to the
white dwarf cooling curve.  Thus, a gap should appear in a CMD of
post-HB stars.  We show this gap as a shaded region in 
Figures~\figmacmd\ and \figmbcmd; it is
obvious in population simulations (cf.\ \S\ref{secehbcmd} and 
Figure~\figsim). Unfortunately, our
CMDs do not reach deep enough to detect this gap; most of the stars
below the thick black line in the figures are either spurious sources
or have large statistical uncertainties, thus blurring any gap if present.

Because these evolved stars spend the vast majority of their time on the
HB before proceeding to their later phases, we 
are resolving only a small fraction of the UV flux in these CMDs
(as discussed in \S\ref{seciue}).  For every star we detect
in the relatively rapid post-HB phases, there are many more sitting
on the horizontal branch.  Until a CMD extends to the HB, most of the
UV flux in these galaxies will remain unresolved.

Main sequence stars or blue stragglers are unlikely contributors to the
population shown in our CMDs.  In Figure~\figms, we show 
the M~31 CMD, but now plot the main sequence instead of post-HB stars.
The zero-age main sequence (ZAMS) is computed from solar-metallicity tracks of
Bressan et al.\ (1993\markcite{B93}), and is labeled with the masses
of the stars at selected points.  Although the main sequence passes through
the population of detected stars, our detected stars do not appear to be
clumping around the ZAMS.  This is in line with expectations, 
because stars of such high mass would be extremely young; the approximate
turnoff ages corresponding to each of the masses in the diagram are:
25 Myr (9 M$_{\odot}$), 40 Myr (7 M$_{\odot}$), 
50 Myr (6 M$_{\odot}$), 80 Myr (5 M$_{\odot}$), 125 Myr (4 M$_{\odot}$), and 
250 Myr (3 M$_{\odot}$).  Although the stars near the detection limit
could possibly be 250 Myr old main sequence stars, it seems implausible
that star formation stopped so recently and abruptly in M~31 and M~32.
Blue stragglers, the possible result of binary mergers, could conceivably
contribute at the detection limits, because a 3~M$_{\odot}$ blue straggler
could be the product of two 1.5~M$_{\odot}$ stars of 2 Gyr age, but 
brighter than these limits their contribution becomes less
plausible.

Figures \figmacmd\ and \figmbcmd\ show what may be a few 
fading PAGB stars along the left hand side of the diagram.  Such
stars may show significant emission lines from any surrounding planetary
nebulae, and we note that the stellar spectra used to trace the tracks in the 
CMDs do not account for such emission lines.  The presence of strong lines 
would tend to move the stars to the left (blue) and up (brighter) in the CMDs,
because the strongest lines fall well within the F175W bandpass.  For
example, in the well-studied planetary nebula NGC 7662, the line strengths of 
\ion{C}{4}$\lambda$1549  and \ion{C}{3}]$\lambda$1908 relative to 
[\ion{O}{3}]$\lambda$5007 are observed to be 0.553 and 0.289, respectively
(Osterbrock 1989\markcite{O89}).  Ciardullo et al.\ (1989\markcite{C89})
measured [\ion{O}{3}]$\lambda$5007 for planetary nebulae in the M~31 bulge 
to have a range of $20.31 \leq m_{5007} \leq 24.53$ mag, translating to 
$ 2.40\times 10^{-14} \geq F_{5007} \geq 4.92\times 10^{-16}$ erg cm$^{-2}$ 
s$^{-1}$.  If for the purposes of demonstration we take the 
brightest [\ion{O}{3}]
flux, compute the intensities in \ion{C}{3}] and \ion{C}{4} (accounting
for the reddening in M~31), and run these fluxes through SYNPHOT,
we find that a T$_{\rm eff}$~=~108000~K PAGB star (0.633 M$_{\odot}$) 
would move significantly
in the M~31 CMD.  Without the emission lines, $m_{F175W}$~=~23.10~mag and
$m_{F175W} - m_{F275W} = -1.29$~mag, but with the emission lines, these would
change to $m_{F175W}$~=~21.22~mag and $m_{F175W} - m_{F275W} = -3.03$~mag.

As was the case with the luminosity functions in \S\ref{seclf}, there do not 
appear to be any dramatic differences between the evolved stellar populations 
of M~31 and M~32 seen in our CMDs.  The stars are for the most part
evenly distributed among the post-HB tracks in both galaxies.  
The only significant difference (other than the larger number of stars
seen in the M~31 CMD) might be the appearance of a greater ``red'' post-HB 
population in M~31.

The statistical uncertainty for the magnitudes of most stars in our
CMDs is large enough to cause considerable ``blurring'' of the CMDs.
For this reason alone, it would not be feasible to fit the
mass distribution for the horizontal branches of M~31 and M~32.   However,
the many systematic uncertainties also 
prevent a more precise characterization at 
this time.   The dominant uncertainty is the reddening applied to the tracks 
(see above).  The stars would shift 0.24~mag to the right or upward 
in the CMDs if the F275W filter or both filters are not correctly calibrated 
(see \S\ref{secf275}).  There are
theoretical uncertainties in the evolutionary tracks themselves.  For
these reasons, we must limit the conclusions we draw regarding
these populations.  The stars shown in our CMDs can be characterized,
as a whole and with considerable certainty, to be evolving from the EHB,
but we cannot use the distribution of stars in the CMDs to determine
the distribution of mass for the horizontal branch progenitors.

\parbox{3.0in}{\epsfxsize=3.0in \hspace*{0.1in} \epsfbox{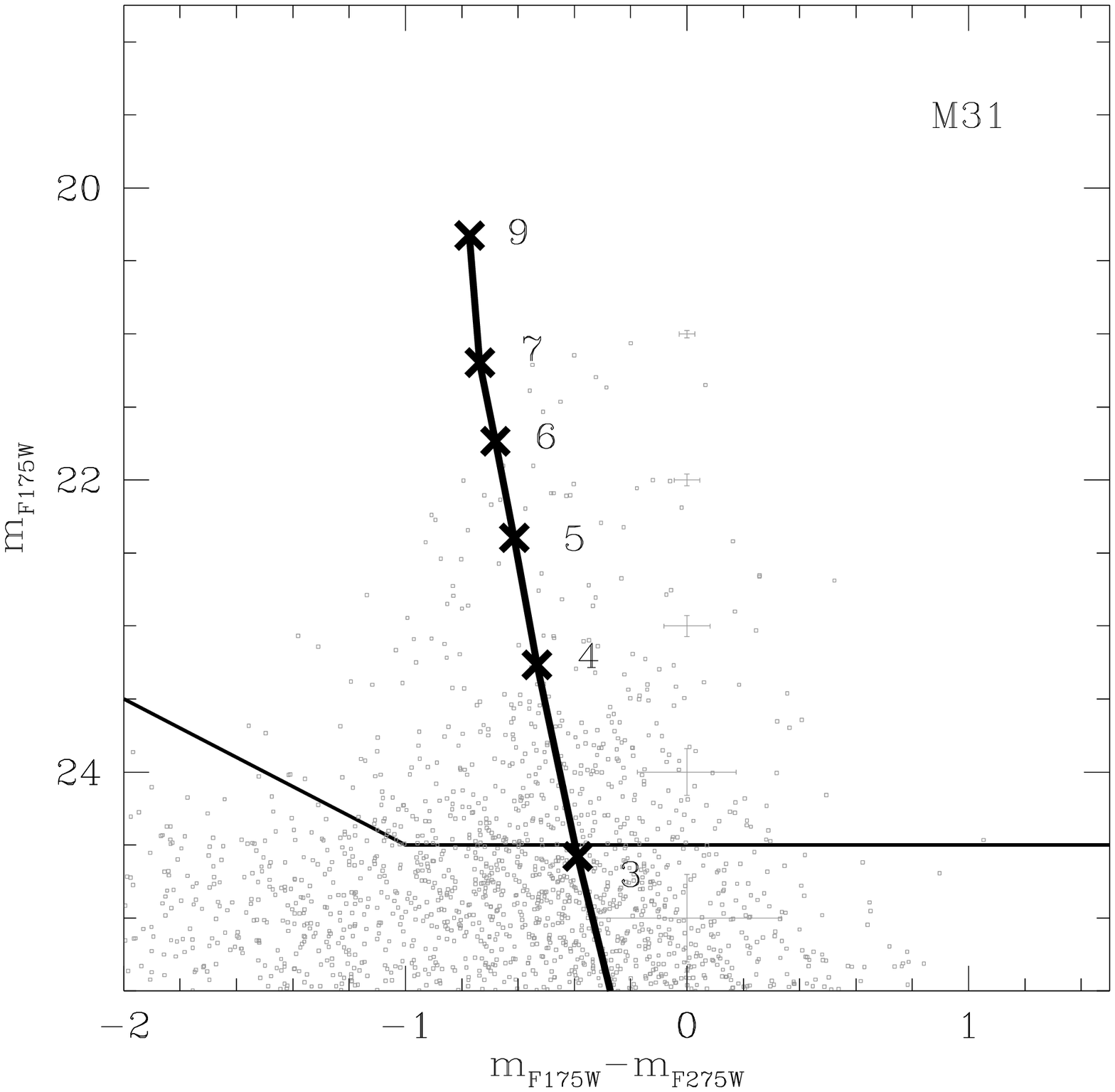}}
\centerline{\parbox{3.0in}{\small {\sc Fig.~\figms--}
The CMD for M~31 (cf.\ Fig.~\figmacmd), but instead of post-HB tracks,
the zero-age main sequence (ZAMS) is plotted for a range of masses
appropriate to the luminosity and temperature of this population,
assuming solar-metallicity and a reddening of $E(B-V)=0.11$~mag.
The masses of the ZAMS stars are labeled in solar units.  Note that only
very recent star formation could populate this portion of the ZAMS.
The stars do not appear to be clumping near the ZAMS (or along
any line shifted with respect to the ZAMS), even where the
photometric errors are small.
}}\vspace{0.2in}
\addtocounter{figure}{1}

\section{Modelling}

\subsection{EHB Luminosity Function}
\label{secehblf}

From the spectral-synthesis results of Ferguson \& Davidsen 
(1993\markcite{FD93}) and Brown et al.\ (1997\markcite{B97}), 
we expect that the LFs are dominated by
post-EHB stars, i.e.,  AGB-Manqu$\acute{\rm e}$ stars and 
PEAGB stars (Greggio \& Renzini 
1990\markcite{GR90}).  Due to the limitations imposed
by the observational uncertainties, detailed fitting of the EHB mass
distribution does not seem warranted.  Furthermore, there exist no a priori
predictions of the HB mass distribution.  However, we can explore the 
simple test case of a flat distribution of mass (i.e., a constant number 
of stars per year per unit EHB star mass joining the zero-age HB), and 
compare to the {\it FOC} data, in order to see if the populations 
characterized in the data are consistent with a population of EHB stars.
Although the EHB tracks appear to be populated (see \S\ref{seccmd}),
a simulation properly accounts for the timescales in the evolutionary
tracks; slower phases should be more populated than faster phases of
post-HB evolution, and such subtleties are not obvious to the eye
in a CMD.

A flat distribution of mass on the EHB reproduces the measured
luminosity functions reasonably well.  In Figure~\figehblf\ we 
show the LFs we obtain from a population evolving from a uniform
distribution of mass on the extreme horizontal branch
(0.471--0.530 M$_{\odot}$), scaled so that the number of stars above
our detection limits matches that seen in M~31.  For this
simulation, we use the evolutionary tracks of Dorman et 
al.\ (1993\markcite{DRO93}); the magnitudes are derived using
SYNPHOT with Kurucz synthetic spectra.  The top panels show
the LFs from the test population as we would measure them all the way down 
to the horizontal 

\begin{figure*}[t]
\centerline{\parbox{5in}{\epsfxsize=5in \epsfbox{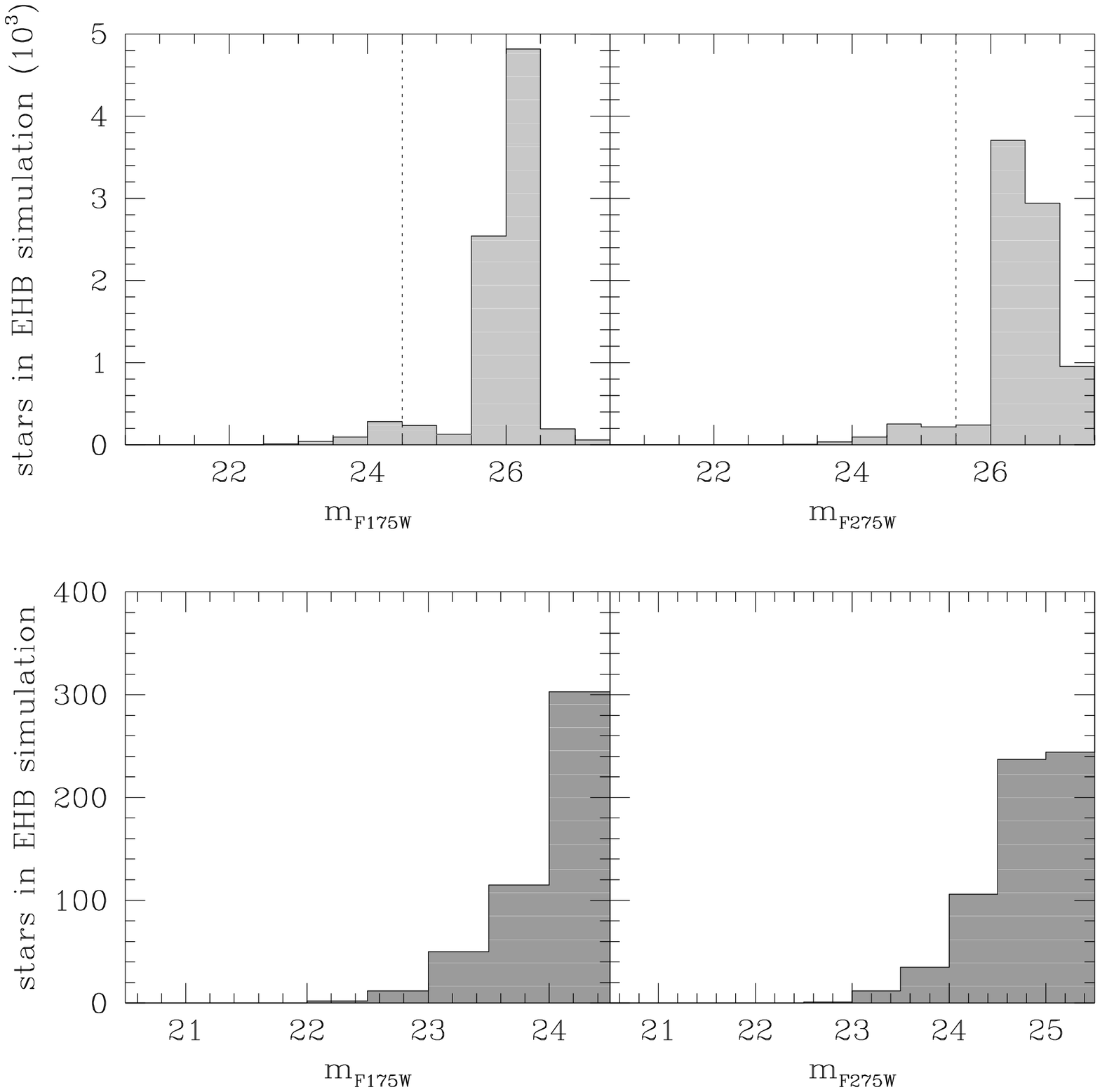}}}
\caption{
The luminosity function for stars entering EHB tracks of mass
0.471--0.530~M$_{\odot}$, 
normalized to reproduce the number of stars found above
the detection limits in the M~31 images.
In the top panels, the luminosity functions do not account for photometric
statistical errors, and are shown all the way down to the HB (dotted lines
denote our actual limiting magnitude in each band).
In the bottom panels, the LFs include photometric errors, and are
shown on the same scale as Figure~\figlf.  Although these LFs
agree with Figure~\figlf\ better than the PAGB LFs in 
Figure~\figpagblf, the actual F275W LF in M~31 includes more faint
red stars than predicted by such a simple model.
}
\end{figure*}

\noindent
branch without photometric uncertainties, and the bottom 
panels show the LFs on the same horizontal scale as the measured LFs in
Figure~\figlf, including the effects of photometric uncertainties.  
The major difference between the simulated LFs in Figure~\figehblf\ and 
the actual M~31 LFs in Figure~\figlf\ is that the measured F275W LF 
peaks higher than the measured F175W LF,
whereas the simulated LFs peak at approximately the same height in each
band.  The measured LFs thus show a redder population than the simulated
EHB LFs.   However, the simulated flat EHB population in 
Figure~\figehblf\ shows much better agreement with the measured LFs 
(Figure~\figlf) than a PAGB-dominated LF (Figure~\figpagblf).

\subsection{EHB Color Magnitude Diagram}
\label{secehbcmd}

For comparison to the actual CMDs (Figures~\figmacmd\ and 
\figmbcmd), we can simulate the CMD we would expect for this
flat EHB mass distribution, again using the evolutionary tracks for these 
stars (Dorman et al.\ 1993\markcite{DRO93}).  The simulation is shown in
Figure~\figsim.  Although they comprise the majority of the population
below the detection limits, PAGB stars are not included in the simulation, 
due to uncertainties along the AGB and in the thermally pulsing phase.  In the
simulation, the total number of stars brighter
than our detection limits in each band matches the number of stars detected
above these limits in M~31.  The top panel shows the simulation without
statistical errors; in the bottom panel, random noise in the colors and 
magnitudes has been added to the simulation, assuming the same data quality
seen in the M~31 observation.  We note that the bottom panel does not take 
into account the increase in spurious sources and decrease in completeness 
as one approaches our detection limits.  The bottom panel in the simulation 
appears similar to the CMD seen in M~31 (Figure~\figmacmd), 
given that the simulation is meant to demonstrate a zeroth order model as an
example.  

A quantitative comparison of the simulated population
(Figure~\figsim b) to the M~31 population (Figure~\figmacmd)
is shown in Figure~\figdens.  The figure partitions the CMDs into
regions that span 0.5 mag in color and 1 mag in luminosity; such partitioning
is appropriate given systematic errors smaller than 0.3 mag.  In each
region we have denoted the number of stars in the M~31 CMD (bold) and the 
number of stars in the simulation (italics).  It is clear that overall, 
the flat distribution
of EHB stars shown in the simulation produces a population that is highly
concentrated toward the lower central regions of the CMD, as seen in M~31.
The M~31 CMD shows a few more red stars along the right hand side, and a few
less blue stars along the left hand side.  The significant difference,
however, is that the M~31 CMD shows considerably more stars brighter
than $m_{F175W}$~=~22.5 mag, as compared to the simulation.  The difference
is large enough to imply a contribution to the population in M~31
that is lacking in the simulation.  

\parbox{5.0in}{\epsfxsize=5.0in \hspace*{0.2in} \epsfbox{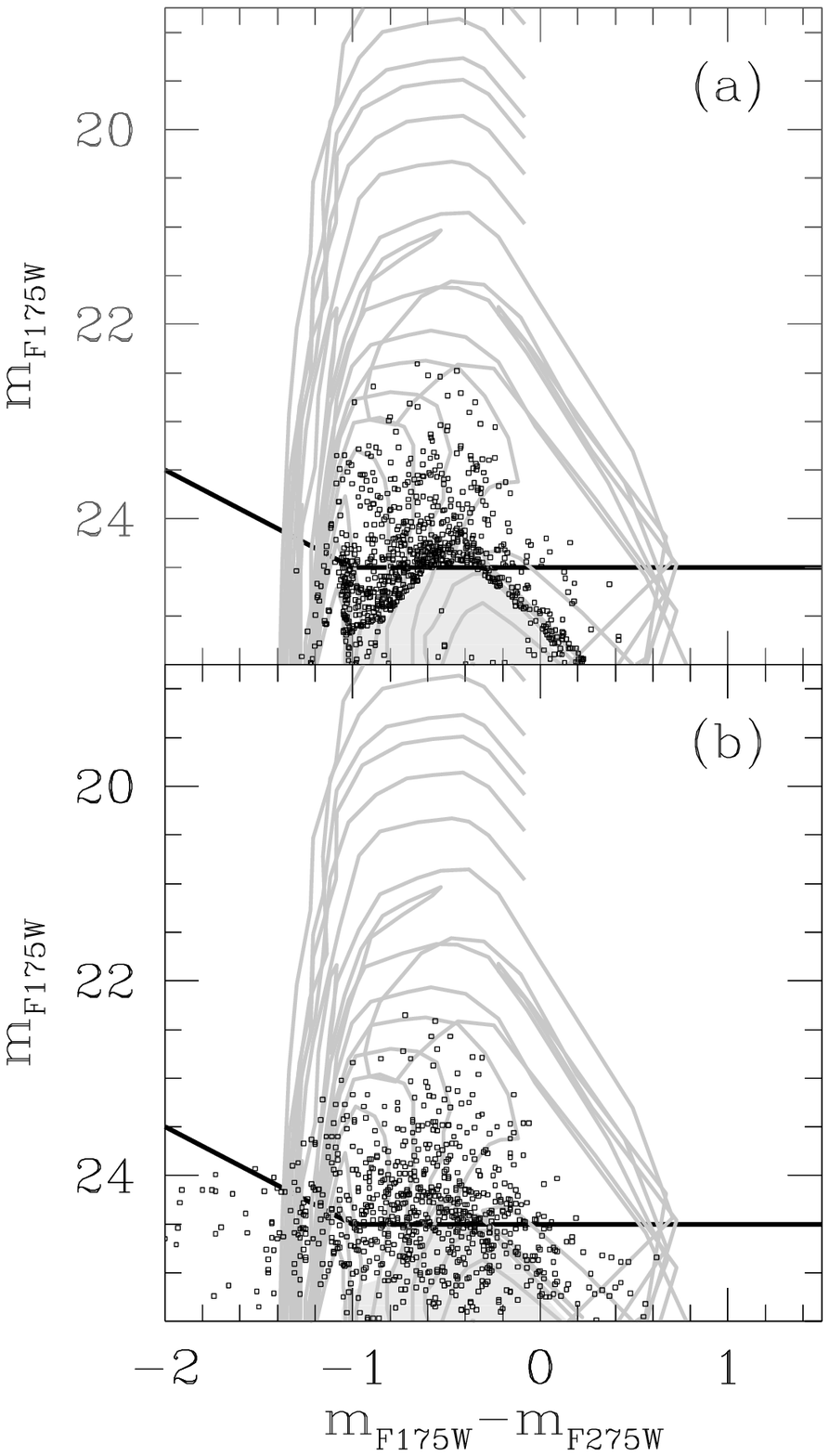}}
\centerline{\parbox{3.0in}{\small {\sc Fig.~\figsim--}
This simulated CMD results from a 
uniformly populated EHB (with ZAHB mass of
0.471--0.530 M$_{\odot}$), but no contribution from the red HB that gives
rise to the PAGB stars.  The number of stars in the simulation above
the detection limits (black line) matches the same number detected in
M~31 (Fig.~\figmacmd).  The top panel shows the simulation without
noise (note the avoidance of the shaded region); the bottom panel 
reflects the statistical errors found in the M~31 data.
}}\vspace{0.2in}
\addtocounter{figure}{1}

What are these brighter stars?  
The simplest way to populate that portion of the CMD is with 
post-early-AGB stars.  Such stars
evolve from the EHB along the AGB, but leave the AGB before
the thermally pulsing phase.  On the zero-age horizontal branch,
they are cooler and more massive than the stars that evolve into
AGB-Manqu$\acute{\rm e}$ stars. PEAGB stars are likely to originate as stars
with masses of 0.51--0.55 $M_{\odot}$ on the ZAHB. While such stars are present
in our simple model, the are evidently not present in sufficient numbers.

However, we find that it is not possible to increase the number of
PEAGB stars without considerably worsening the agreement with the
far-UV spectrum. To match the counts in the bright portion of the CMD,
the population of PEAGB stars would have to be increased by a factor 
of $\sim$100. However this additional population would make 
the M~31 integrated spectrum three times brighter and
significantly cooler than actually observed by {\it HUT}.  
Thus, we consider it unlikely that the very brightest stars in the 
M~31 CMD are PEAGB stars. 
It also seems unlikely that these stars are merely the result of crowding,
given the modest number of resolved stars in our M~31 images, and
the existence of such stars even near the edges of our fields.
Thus the nature of the $m_{F175W} < 22.5$ population is a bit of a
puzzle. Testing other hypotheses (e.g.\ reddened PAGB stars, binary
AGB-Manqu$\acute{\rm e}$ stars, or accreting 
white dwarfs) is beyond the scope of
this paper, but appears to be an interesting avenue for further 
investigation.

\parbox{3.0in}{\epsfxsize=3.0in  \epsfbox{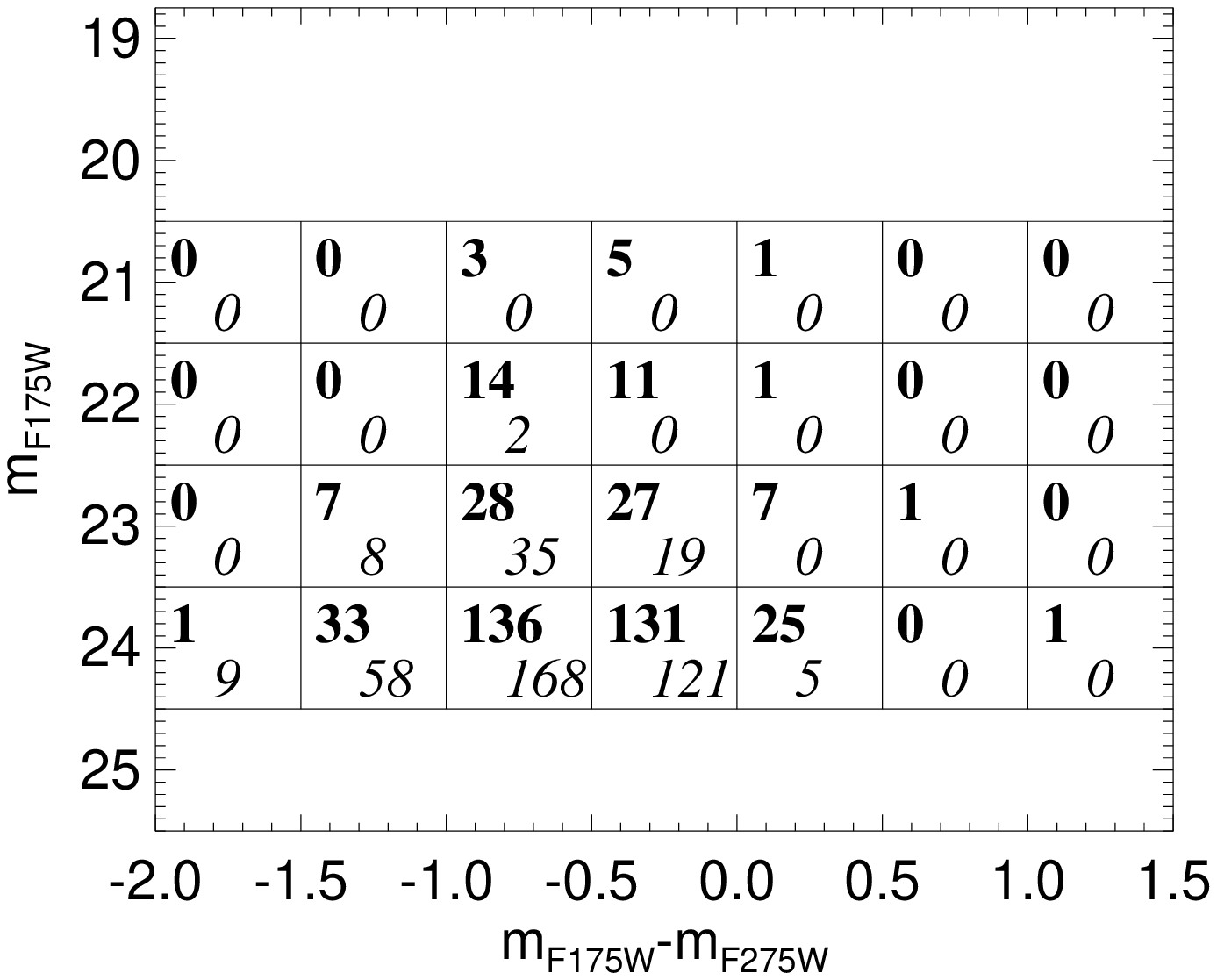}}
\centerline{\parbox{3.0in}{\small {\sc Fig.~\figdens--}
A quantitative comparison of the CMDs shown in Figures~\figmacmd\ and
\figsim b shows good agreement in general but notable discrepancies.
The number of stars per 0.5~mag $\times$ 1.0~mag region in the M~31 CMD
is shown in bold, while that number in the EHB simulation is shown in italics.
The M~31 population and the simulated EHB population both show a strong
concentration toward the faint center.  However, there are $\sim$30 stars
at $m_{F175W} \geq 22.5$~mag in the M~31 CMD that are not present in the
simulated CMD.
}}\vspace{0.2in}
\addtocounter{figure}{1}

\subsection{Comparison to Expectations from {\it IUE} and {\it HUT}}
\label{seciue}

How much of the {\it IUE} flux do we resolve into stars?  Does the {\it FOC}
flux (both diffuse and stellar) agree with the expectations from 
{\it HUT} and {\it IUE}?  In \S\ref{secf275wfpc}, we compared 
the {\it IUE}+Optical flux to the {\it FOC} and {\it WFPC2} flux within 
1.4$\arcsec$ of the galaxy centers, in order to minimize uncertainties in 
the sky and dark count background while investigating any systematic errors 
in the {\it FOC} calibration.  Here, we
compare the {\it FOC} flux in the full {\it IUE} aperture to expectations 
from the {\it IUE}+Optical spectrum, and compare the stellar light to
expectations from both {\it IUE} and {\it HUT}.

Because the $10\times20\arcsec$ oval {\it IUE} aperture is larger than the 
14$\times14\arcsec$ {\it FOC} field, we have defined two regions 
(see Figures \figmaschem\ and \figmbschem) 
in the {\it FOC} field for comparison to the {\it IUE} flux.  Region ``A'' is
centered on the nucleus in each galaxy, and region ``B''  runs from ``A''
to a distance of 10$\arcsec$ from the galaxy center.  The regions are
defined such that A+2B is equivalent to the {\it IUE} aperture, given 
approximate symmetry in the galaxies about their centers.  

The net {\it FOC} count rate within the {\it IUE} aperture is shown in 
Table~\ref{tabiue} (row 4), and is the result of subtracting the 
expected sky count rate (row 3) and expected dark count rate (row 2) from the
gross count rate (row 1).  We again assume a high sky background,
although unlike our calculations in \S\ref{seccompspur} and 

\newpage

\parbox{5.25in}{\epsfxsize=5.25in \epsfbox{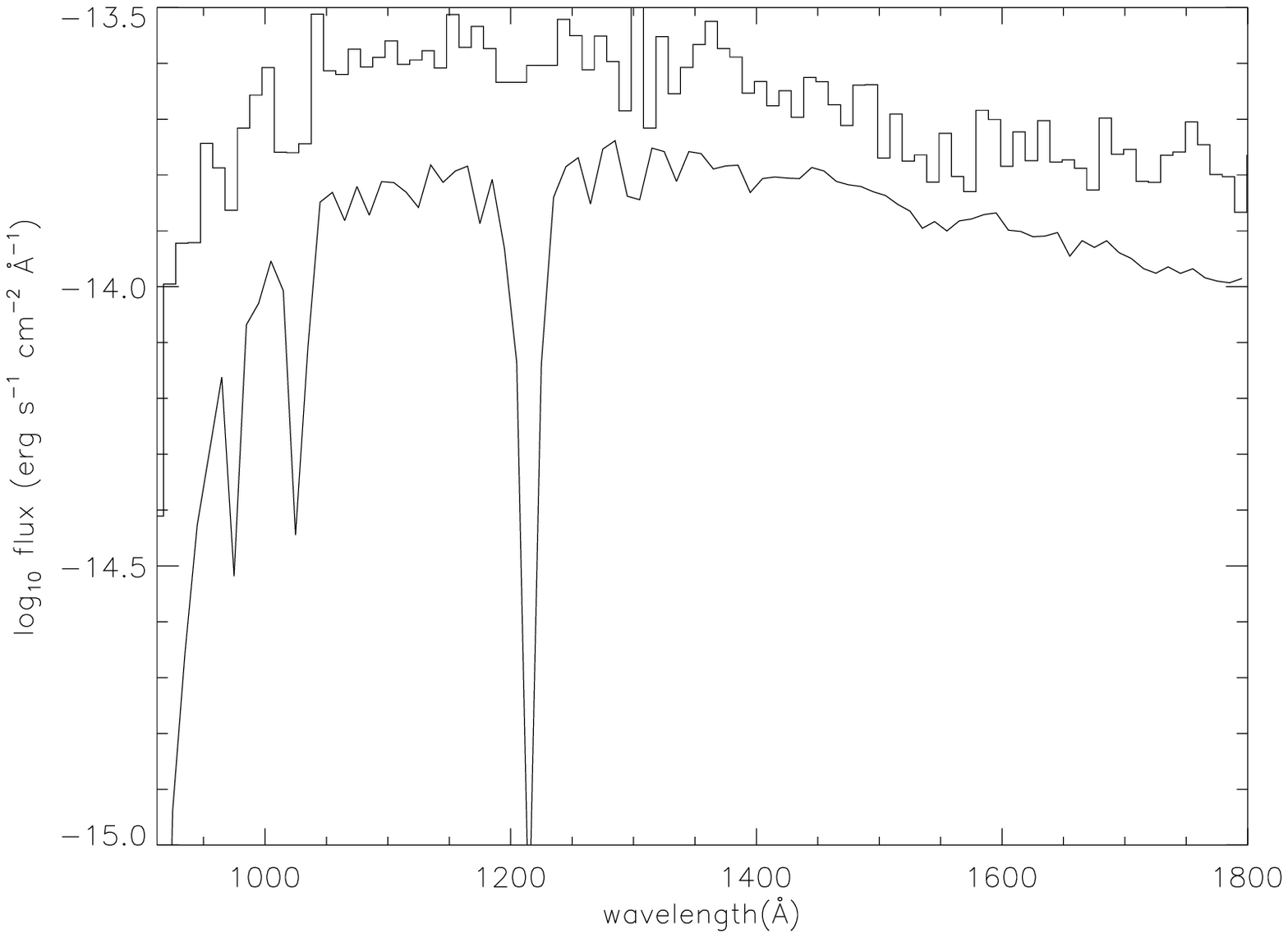}}
\parbox{6.5in}{
\small {\sc Fig.~\figmaspec--}
The {\it HUT} spectrum of M~31 (histogram; binned to 10~\.{A}) shown with
the spectrum we would obtain from a two-component population of 
evolved stars: a flat distribution of EHB stars, and a population of
0.633~M$_{\odot}$ PAGB stars.  The number of EHB stars was chosen to
match the number above the detection limits in the M~31 images,
and the number of PAGB stars was chosen to match that allowed
under fuel consumption constraints (accounting for the small contribution
from the EHB stars); the entire model was then multiplied by 3.59 to
account for the aperture difference between the {\it FOC} stellar sample and
the {\it HUT} $9.4 \times 116 \arcsec$ slit.
}
\vspace*{0.25in}
\addtocounter{figure}{1}

\parbox{5.25in}{\epsfxsize=5.25in \epsfbox{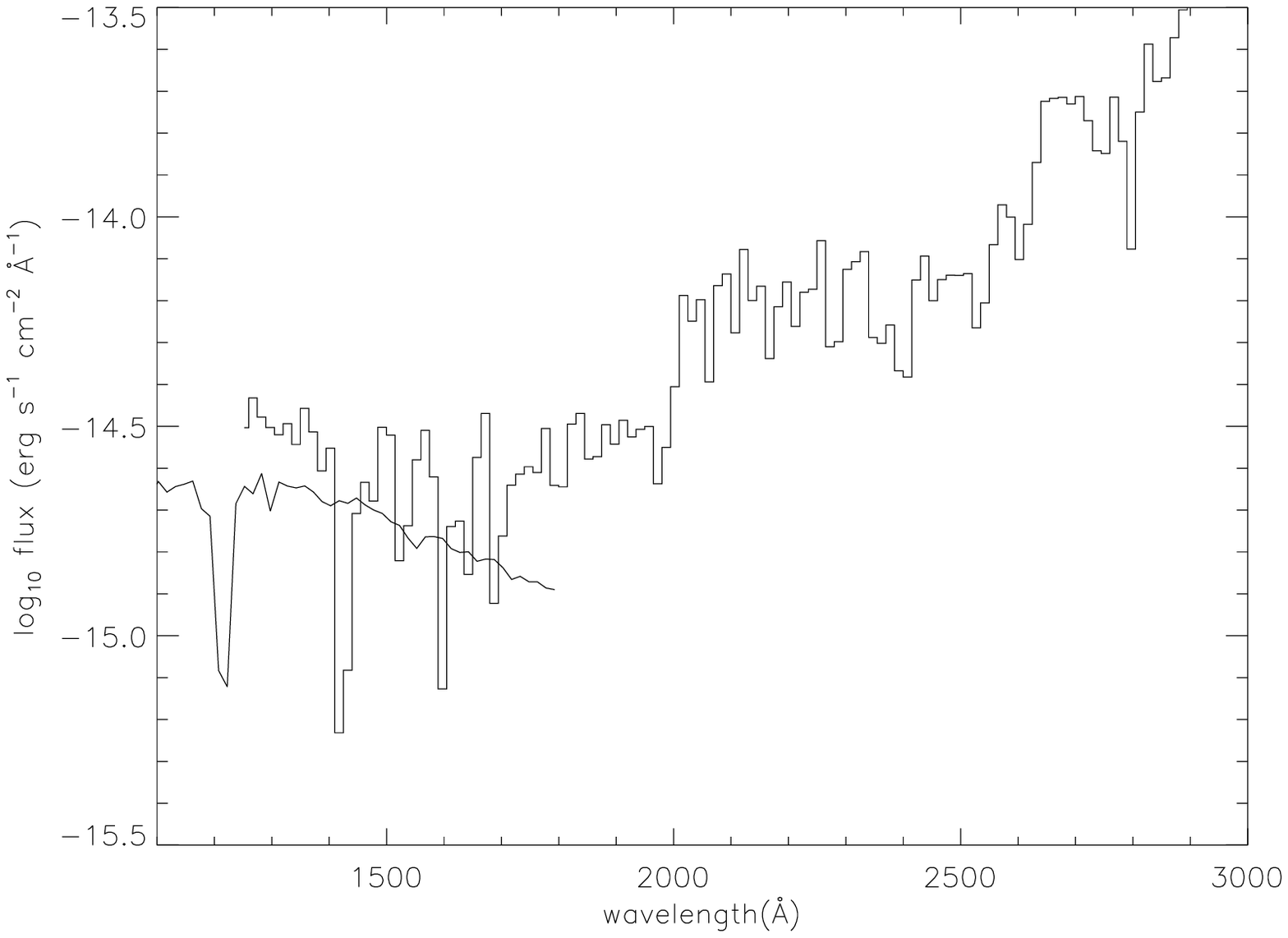}}
\parbox{6.5in}{
\small {\sc Fig.~\figmbspec--}
The {\it IUE} spectrum of M~32 (histogram; binned to 15~\.{A}) shown with
the spectrum we would obtain from a two-component population of 
evolved stars: a flat distribution of EHB stars, and a population of
0.633~M$_{\odot}$ PAGB stars.  The number of EHB stars was chosen to
match the number found 
above the detection limits in the M~32 images, within the
region ``A'' plus twice that number found in region ``B'' 
(cf.\ Figure~\figmbschem), thus accounting for the difference
between the {\it FOC} and {\it IUE} apertures.
The number of PAGB stars was chosen to match that allowed
under fuel consumption constraints (accounting for the small contribution
from the EHB stars).  
}
\addtocounter{figure}{1}

\clearpage

\noindent
Table~\ref{tabcompspur}, 
the sky and dark background contribute significantly
to the flux within this large area.  Running the composite
{\it IUE}+Optical spectrum through the IRAF package SYNPHOT gives the 
predicted flux within the {\it IUE} aperture (row 6), and a comparison (row 8)
of this expectation with our measurement shows that our {\it FOC} 
F175W images detect more flux than predicted from {\it IUE}, while
the F275W images detect less flux than predicted. 
Given the uncertainty in the dark and sky background
for the {\it FOC}, we can say that the flux measured by the 
{\it FOC} cameras roughly agrees with the predictions made from the 
{\it IUE}+Optical composite spectrum.

To compare to the flux in resolved sources, we limit ourselves to those
stars that are above our limiting magnitudes (i.e., $m_{F275W}<25.5$ mag and
$m_{F175W}<24.5$ mag).  We compare the flux from the stars found in the 
{\it IUE} ``aperture''
(again regions A+2B) to the UV flux seen by {\it IUE}.  We define the UV
flux in the F175W filter to be that flux shortward of 2400~\.{A}, and
the UV flux in the F275W filter to be that flux shortward of 3400~\.{A};
flux longward of these wavelengths is considered ``red leak'' in the
filters.  As the numbers in Table~\ref{tabiue} demonstrate, the F175W filter
suffers from a very high red leak, while the F275W filter shows a
moderate contribution from red leak photons.  The high percentage of
red leak in these calculations is due to the fact that both galaxies are
very red and emit
much more light at longer wavelengths than in the UV when integrating
over a synthesized $10 \times 20 \arcsec$ aperture; we note that the 
percentage of red leak in the light from hot stars is far less important.  
For example, for {\it FOC} photometry of a 20000~K star in M~31 or M~32, 
the red leak contribution to the measured light is less than 12\% in 
F175W and less than 3\% in F275W.  If we assume 
that the stars above our limiting magnitudes emit the bulk of their flux
in the UV, it is reasonable to compare (row 9)
the counts found in these stars to the UV count rate predicted from
{\it IUE}.   The UV flux from our resolved hot stars only
accounts for a small fraction ($<$20\%) of the UV light observed with 
{\it IUE}.  This result is understood if we trace these post-HB stars back to
their HB progenitors.

As we did in Sections \ref{secehblf} and \ref{secehbcmd}, we again use a flat 
distribution of stars on the EHB and its post-HB descendents, and then scale 
this population so as to reproduce the number of stars found above our 
detection limits in M~31 and M~32.  Note that many more stars in this model 
are below the detection limits than above (see Figure~\figehblf).
We then add a 0.633~M$_{\odot}$ PAGB component,
with the number of stars scaled to match the limits allowed
under the fuel consumption theorem (taking into account the
small contribution from the EHB stars).  Because the UV-bright phase of
these PAGB stars is so short-lived, only a dozen or so would appear
in each of the M~31 and M~32 images (see Figure~\figpagblf).  
We then use the spectral library of Brown, Ferguson, \& Davidsen 
(1996\markcite{BFD96}) to compute the integrated spectra of these populations, 
and compare to the actual far-UV spectra of M~31 and M~32 as measured by 
{\it HUT} and {\it IUE} respectively.  These comparisons are shown in Figures
\figmaspec\ and \figmbspec.  Note that the model
spectrum in each figure has been rescaled to match the flux
expected in the {\it HUT} and {\it IUE} apertures, but has not
been renormalized otherwise.  The disagreements between the model and the
observed spectra are thus real and due either to a failure of this
aperture correction to match the true UV profile of the galaxies,
or a missing UV population in the model.

The figures demonstrate that most of the unresolved flux in M~31 and
M~32 can be explained, once we trace back the bright post-HB population
detected in the FOC to its HB progenitors.  Given the large uncertainty
in the aperture matching between the {\it FOC}, {\it HUT}, and {\it IUE},
the agreement between the model spectrum and the far-UV data is good.
For M~31, a model population that reproduces the number of
detected stars and the UV light has 98\% of its main sequence stars
evolving into PAGB stars, 
and only 2\% into EHB stars (with a flat mass distribution).  For M~32, a model
that reproduces the number of detected stars and the UV light has
99.5\% of its main sequence stars evolving into PAGB stars and 
0.5\% into EHB stars.   The far slower evolution of the EHB stars more
than compensates for the smaller number of their progenitors relative
to those for the PAGB stars.  The smaller fraction
of resolved UV flux in M~32 is explained by a smaller fraction
of stars following the longer-lived EHB evolutionary paths.
Although the uncertainties are large, the model spectrum 
in each galaxy falls a bit short of those measured by {\it HUT}
and {\it IUE}, and the discrepancy is largest at the shortest
wavelengths.   Very hot EHB stars (and their descendants),
with envelope mass less than 0.002~M$_{\odot}$, could be present and 
make significant contributions to the far-UV flux in these
galaxies.  These stars are just below are detection limits, even
in their brightest phases, as we discuss in \S\ref{seccmd}.

\subsection{PAGB Circumstellar Extinction}
\label{secext}

We demonstrated in \S\ref{seclf} that the luminosity functions in M~31 and
M~32 do not appear to be dominated by PAGB stars.  Nor do the PAGB
tracks in the color magnitude diagrams (\S\ref{seccmd} and Figures
\figmacmd\ and \figmbcmd) appear to be populated.  
Our analysis showed that a dominant population of low-mass 
(0.569 M$_{\odot}$) PAGB stars would produce many detections in our {\it FOC}
images, because of their relatively slow evolution through the earlier 
bright phases, while a dominant population of 
intermediate-mass PAGB stars (0.633 M$_{\odot}$) stars would produce a 
negligible number of detections due to more rapid evolution.  Given
a population that is dominated by PAGB stars (see \S\ref{seciue}), our data
implies that either the PAGB stars are of intermediate mass (or greater),
or that they are low-mass stars hidden by circumstellar extinction
during their early bright phases.  In this section we explore the latter case.

K$\ddot{\rm a}$ufl, Renzini, \& Stanghellini (1993\markcite{KRS93})
derived a simple model for circumstellar extinction in PAGB stars.  This
extinction would be caused by the shell of material expelled on the AGB
during the ``superwind'' phase.  The initially opaque shell would begin
thinning with the end of the superwind phase, as it expands outward from
the underlying star.  K$\ddot{\rm a}$ufl et al.\ computed 
the ``thinning time'' for this
shell to become transparent at H$\alpha$ and Br$\alpha$.  
We can also apply the K$\ddot{\rm a}$ufl et al.\ model to see the
maximum effect of such an expanding shell in the far-UV.  
Of the 7 models
considered by K$\ddot{\rm a}$ufl et al.\ (1993\markcite{KRS93}), the one
with the strongest and most prolonged extinction in the far-UV is Model 1,
having a superwind mass loss rate of 10$^{-3}$ M$_{\odot}$ yr$^{-1}$,
grain size 0.1 $\mu$m, expansion velocity 10 km s$^{-1}$, and superwind
duration 10$^3$ yr.  The absorption coefficient for 0.1 $\mu$m grains
is respectively 1.348 and 0.413 at 1750 and 2750 \.{A} Draine \& 
Lee (1984\markcite{DL84}).  The time at which the superwind phase ends
is uncertain, but is thought to occur somewhere around an effective temperature
of 5000--7000~K, on or near the AGB (K$\ddot{\rm a}$ufl et 
al.\ 1993\markcite{KRS93}; Vassiliadis \& Wood 1994\markcite{VW94}).
Because we are investigating the maximum effect of such a shell, we shall
chose to quench the superwind immediately prior to the start of
the PAGB tracks (cf.\ Figures~\figmacmd\ and \figmbcmd),
at 10000~K, thus prolonging the thinning as far possible into the UV-bright 
phase of the PAGB evolution. 

We demonstrate the effect of such circumstellar (CS) extinction on the
low-mass (0.569 M$_{\odot}$) PAGB track.  We found in \S\ref{seclf} that
if the PAGB stars in the M~31 and M~32 cores were following such low-mass
tracks, they should be detectable in the {\it FOC} in significant numbers,
as long as CS extinction was not significant.  Figure~\figext\ shows
$m_{F175W}$ and $m_{F275W}$ as a function of time for such stars,
with and without CS extinction.  It takes approximately 10000 years for
the circumstellar shell to become optically thin at 1750 \.{A}, and
4000 years to become optically thin at 2750 \.{A}.  We can see that the
luminosity functions we measured in the {\it FOC} images would still
show a bright peak well above our detection limits, even with the CS
extinction.  A more realistic model would begin the thinning of the 
shell at time several 10$^3$ yr earlier, and in such a case the extinction
would have a very negligible effect on our data.

\vspace*{0.2in}

\parbox{3.0in}{\epsfxsize=3.0in \hspace*{0.1in} \epsfbox{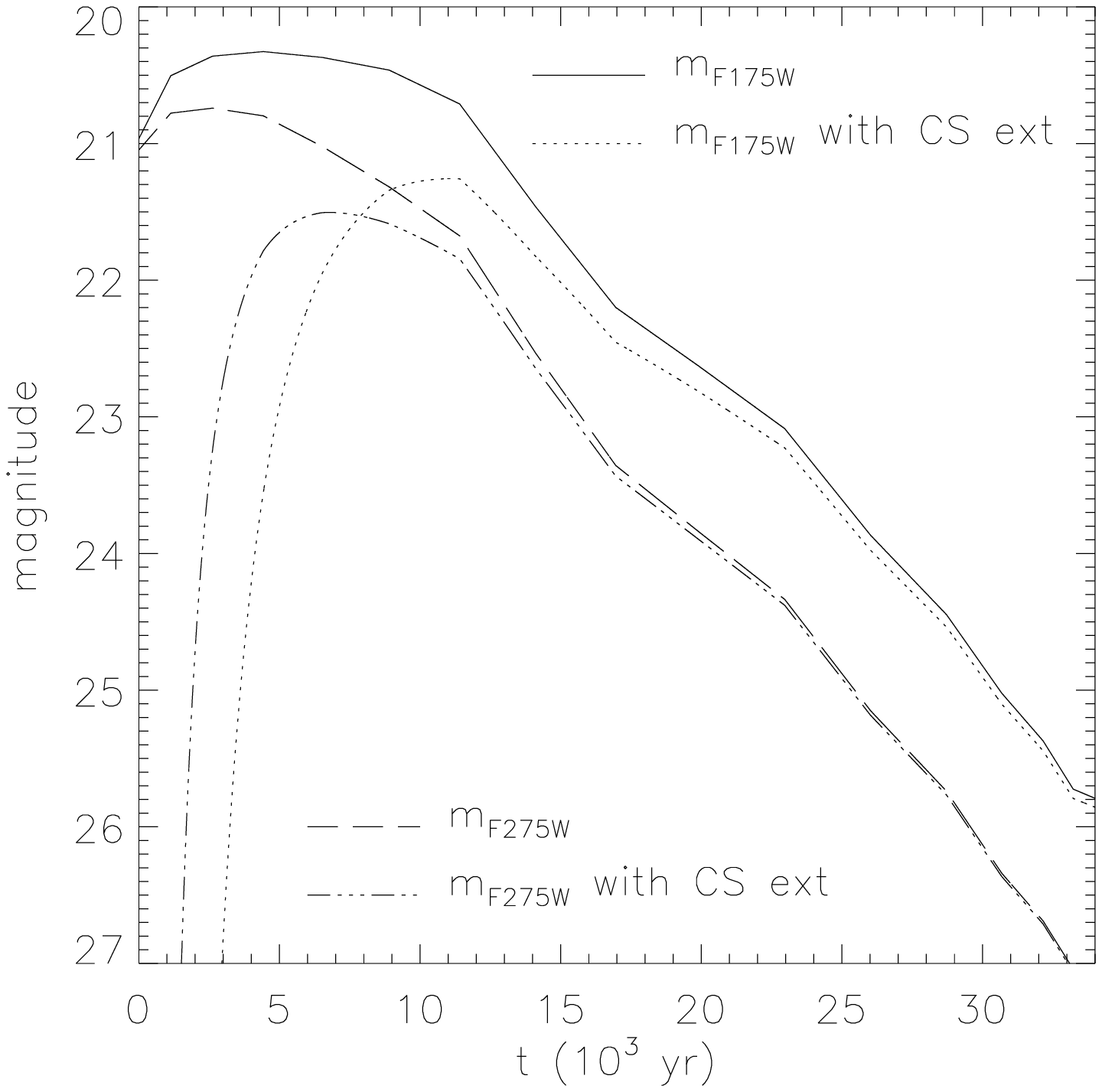}}

\vspace*{0.4in}

\centerline{\parbox{3.0in}{\small {\sc Fig.~\figext--}
The evolution of a low-mass (0.569 M$_{\odot}$) PAGB track as observed
with and without circumstellar extinction.  The CS extinction model
has been chosen to maximize the effect in the far-UV, but it would still
not significantly affect the detection of such low-mass PAGB stars in the
{\it FOC} field.  Intermediate-mass PAGB stars are thus more likely
to dominate the populations in M~31 and M~32.
}}\vspace{0.2in}

\section{DISCUSSION}
\label{secdis}

Since the discovery of the UV upturn phenomenon, a great deal of research has
tried to characterize the stellar population producing the UV light
in spiral bulges and ellipticals.  The range of candidates has included
young massive stars, hot horizontal branch stars, planetary nebula
nuclei (PAGB stars), and several binary scenarios
(see Greggio \& Renzini 1990\markcite{GR90} for a complete review).
Earlier work relied primarily upon {\it IUE} and 
{\it HUT} spectra of composite stellar
populations, and pre-COSTAR {\it FOC} observations (King 
et al.\ 1992\markcite{K92}; Bertola et al.\ 1995\markcite{BBB95}) of the
brightest stars.
Our current understanding of the populations in ellipticals and spiral
bulges predicts that a significant population of hot HB and post-HB
stars should be evolving from the extreme horizontal branch and 
following either AGB-Manqu$\acute{\rm e}$ or post-early-AGB evolution
(see Brown et al.\ 1997\markcite{B97} and references therein).  
The prediction is mainly based upon fuel consumption constraints that
rule out PAGB stars as the sole UV producers; the PAGB stars
are just too short-lived to efficiently produce the required UV light.

Our {\it FOC} observations confirm the existence
of hot post-EHB stars in the centers of M~31 and M~32,
and also confirm that PAGB stars are not the predominant component
of the UV-bright population.  The existence of
these EHB stars also implies that the horizontal branches in both M~31 and M~32
are at least somewhat extended, because ``red clump'' horizontal branches are
unlikely to produce the stars seen in our color-magnitude diagrams.
Future observations, given the improved UV capabilities of HST with
the installation of {\it STIS}, should
be able to reach the HB and thus characterize the entire evolved population
more directly.  We note that significant systematic uncertainties may exist
in our {\it FOC} photometry, at the 0.3 mag level.  Our results should
be considered with these uncertainties in mind; however, the uncertainties are
not large enough to allow a completely different interpretation of the
detected stellar population.

As discussed in \S\ref{secehbcmd}, we find
a minority population ($\sim$ 10\%) of brighter stars 
that cannot be explained by canonical post-HB evolutionary tracks.
Although PEAGB or very low-mass PAGB stars evolve through this region
of the CMD, their evolution during this phase is too rapid to produce
the 35 stars seen in M~31 while maintaining consistency with fuel
consumption constraints and observed integrated spectra of M~31. 
The nature of these stars remains unexplained as of this writing.

The fraction of light in the resolved UV
population in the center of M~31 and in the center of M~32 is 
consistent with expectations from the {\it HUT} and {\it IUE} spectra
of these galaxies.  The far-UV light in M~31 can be explained by
a main sequence population where 98\% of the stars channel through
intermediate-mass PAGB evolution and
2\% through EHB evolution.  The far-UV light in M~32 can be explained
by a population where 99.5\% of the stars channel through 
intermediate-mass PAGB evolution and
0.5\% through EHB evolution.  
Our simple model populations can account
for most of the far-UV light, although there is room in both galaxies
for a contribution from very low-mass hot 
EHB stars, which would be below our detection limits even in the
AGB-Manqu$\acute{\rm e}$ phase.   

We found that the stellar populations in
M~31 and M~32 do not appear remarkably different in our {\it FOC} UV images.
Fewer stars appear to be entering the UV-efficient EHB paths in M~32,
and this explains both the smaller absolute number of detected stars and
the smaller fraction of resolved UV light.
The similarities of the luminosity functions of M~31 and M~32 were
unexpected, given the dramatically different $m_{1550}-V$ colors of the two
galaxies. This finding suggests that, while the fractional mass in EHB
stars is probably sensitive to the properties of the overall
stellar population, (e.g., metallicity, age, or helium abundance), the
mass distribution on the EHB may not be as sensitive to these
parameters.  Certainly, the differences
can not be investigated without deeper images that can resolve more of
the evolved population.  {\it STIS} observations planned for late 1998 will
reach the HB in M~32 and provide direct evidence for the horizontal branch
distribution.

\acknowledgments

Support for this work was provided by NASA through grant number GO-5435
from the Space Telescope Science Institute, which is operated by the
Association of Universities for Research in Astronomy, Incorporated,
under NASA contract NAS5-26555.
TMB acknowledges support at Johns Hopkins University by NAS 5-27000 and at 
Goddard Space Flight Center by NAS 5-6499D.
The work by SAS at IGPP/LLNL was performed under the 
auspices of the US Department of Energy under contract 
W-7405-ENG-48.  We wish to thank Robert Jedrzejewski and Ivan King for
providing insight into {\it FOC} calibration issues.

\begin{table}[b]
\caption{{\it FOC} Observations and Photometry}
\label{tabobs}
\begin{tabular}{|l||r|r||r|r|}
\tableline \tableline
& \multicolumn{2}{c||} {M~31} & \multicolumn{2}{c|}{M~32} \\
\tableline
R.A.$_{2000}$ & \multicolumn{2}{c||} {$0^h42^m44.74^s$} &  \multicolumn{2}{c|}{$0^h42^m41.53^s$} \\
Dec.$_{2000}$ & \multicolumn{2}{c||} {$41^\circ 16\arcmin 8.04 \arcsec$} &  \multicolumn{2}{c|}{$40^\circ 51\arcmin 51.82\arcsec$} \\
$D$ (kpc) & \multicolumn{2}{c||} {770} &  \multicolumn{2}{c|}{770} \\
$E(B-V)$\tablenotemark{a} (mag) & \multicolumn{2}{c||} {0.11} & \multicolumn{2}{c|}{0.11} \\
$m_{1550}-V$\tablenotemark{b} (mag)& \multicolumn{2}{c||}{3.51} & \multicolumn{2}{c|}{4.50} \\
\tableline
\tableline
& F175W & F275W & F175W & F275W \\
\tableline
EXPTIME (sec) & 19773 & 8390 & 16179 & 8990 \\
$m_o$\tablenotemark{c} & 19.43 & 21.47 & 19.43 & 21.47 \\
PHOTFLAM\tablenotemark{d} & 6.12$\times10^{-17}$ & 9.38$\times10^{-18}$ & 6.12$\times10^{-17}$ & 9.38$\times10^{-18}$ \\
Pivotal $\lambda$ (\.{A}) & 1990 & 2800 & 1980 & 2800 \\
Band width (\.{A}) & 386 & 323 & 386 & 323 \\
PSF FWHM\tablenotemark{e} (pixel) & 4.6 & 4.0 & 4.3 & 3.9 \\
\tableline 
\end{tabular}
\tablenotetext{}{}
\tablenotetext{a}{Ferguson \& Davidsen (1993\markcite{FD93}).}
\tablenotetext{b}{Burstein et al.\ (1988\markcite{B88}).}
\tablenotetext{c}{The magnitude that produces 1 count/sec in the filter.}
\tablenotetext{d}{The flux (erg cm$^{-2}$ s$^{-1}$ \.{A}$^{-1}$) that 
produces 1 count/sec in the filter.}
\tablenotetext{e}{Estimated from PSF fitting in each image.}
\end{table}

%

\begin{table}
{
\caption{Comparison of {\it WFPC2} and {\it FOC} Integrated Photometry}
\label{tabf275}
\begin{tabular}{|l||l|c|c|c|c|c|c|c|}
\tableline \tableline
(1)  &  (2)   &  (3)     &  (4)          &   (5)         &  (6)          &   (7)         &    (8)    &  (9) \\
     &        &          & Gross         & Dark          & Sky           & Net           & Predicted  & \\
     &        & EXPTIME  & count rate    & count rate    & count rate    & count rate    & Count rate  & \\
Galaxy&Filter & (sec)    & (cts s$^{-1}$)\tablenotemark{a}&(cts s$^{-1}$)\tablenotemark{a} &(cts s$^{-1}$)\tablenotemark{a} &(cts s$^{-1}$)\tablenotemark{a} & (cts s$^{-1}$)\tablenotemark{b} & Net/Predicted\\
\tableline
\tableline
M 31 & F175W  &   19773  &  55.5         &   22.0        &  3.0          &  30.5         & 27.4            & 1.11\\
M 31 & F275W  &   8390   &  228          &   22.0        &  7.6          &  198          & 229             & 0.86\\
M 31 & F300W  &   3000   &  61.0         &   0           &  0.8          &  60.2         & 57.0            & 1.06\\
M 31 & F336W  &   460    &  99.3         &   0           &  0.3          &  99.0         & 103             & 0.96\\
M 31 & F555W  &   2200   &  4811         &   0           &  8.7          &  4802         & $\equiv$4802            & $\equiv$1\\
\tableline
M 32 & F175W  &   16179  &  97.9         &   22.0        &  3.0          &  72.9         & 68.0            & 1.07\\
M 32 & F275W  &   8990   &  684          &   22.0        &  7.6          &  654          & 861             & 0.76\\
M 32 & F300W  &   3000   &  230          &   0           &  0.8          &  229          & 206             & 1.11\\
M 32 & F336W  &   1200   &  339          &   0           &  0.3          &  339          & 364             & 0.93\\
M 32 & F555W  &   104    &  21019        &   0           &  17.4         &  21002        & $\equiv$21002   & $\equiv$1\\
\tableline
\end{tabular}
\tablenotetext{}{}
\tablenotetext{a}{Measured count rate through a 2.8$\arcsec$ diameter
circular aperture.}
\tablenotetext{b}{Predicted count rate based upon the {\it IUE} spectrum, 
including red leak and normalized to agree in the visual (F555W).}
}
\end{table}

\begin{table}
{
\caption{Photometry Characteristics}
\label{tabcompspur}
\begin{tabular}{|l||r|r||r|r||r|r||r|r||r|r||r|r|}
\tableline
 & \multicolumn{4}{|c||}{Completeness (\%)\tablenotemark{a}}
 & \multicolumn{4}{c||}{Standard Deviation (mag)\tablenotemark{a}}
 & \multicolumn{4}{c|}{Spurious sources (\%)\tablenotemark{a}} \\
\tableline
& \multicolumn{2}{c||} {M~31} & \multicolumn{2}{c||}{M~32}
& \multicolumn{2}{c||} {M~31} & \multicolumn{2}{c||}{M~32} 
& \multicolumn{2}{c||} {M~31} & \multicolumn{2}{c|}{M~32} \\
\tableline
mag & F175W & F275W & F175W & F275W
 & F175W & F275W & F175W & F275W
 & F175W & F275W & F175W & F275W \\
\tableline
21.0 & 100 & 100 & 100 & 100 & 
0.034 & 0.027 & 0.036 & 0.027 &
 0  & 0  & 0  & 0  \\
22.0 & 100 & 99 & 100 &  100 &
0.051 & 0.030 & 0.060 & 0.041 &
  0  & 0  & 0  & 0  \\
23.0 & 100 & 100 &  98 & 100 &
0.155 & 0.047 & 0.094 & 0.062 &
  0  & 0  & 0  & 0 \\
24.0 & 86  & 100  &  86 & 100 &
0.252 & 0.124 & 0.193 & 0.125 &
  0  & 0  & 7  & 7 \\
25.0 & 31  & 93  &  32 &  95 &
0.274 & 0.224 & 0.426 & 0.216 &
  60 & 1  & 87 & 6 \\
26.0 & 7   & 46  &  11 & 58 & 
1.056 & 0.519 & 0.682 & 0.446 &
 200& 26 & 188& 35 \\
\tableline
\end{tabular}
\tablenotetext{}{}
\tablenotetext{a}{Beyond 1.5$\arcsec$ of the galactic center.}
}
\end{table}

\begin{table}
{
\caption{Comparison of {\it FOC} and {\it IUE} fluxes}
\label{tabiue}
\begin{tabular}{|l||r|r||r|r|}
\tableline \tableline
& \multicolumn{2}{c||} {M~31} & \multicolumn{2}{c|}{M~32} \\
\tableline
& F175W & F275W & F175W & F275W \\
\tableline
(1) Gross count rate (cts s$^{-1}$)\tablenotemark{a}& 1034 & 2694 & 1070 & 3551 \\
(2) Dark count rate (cts s$^{-1}$)\tablenotemark{b}& 637 & 637 & 637 & 637 \\
(3) Sky count rate (cts s$^{-1}$)\tablenotemark{c} & 87 & 219 & 87 & 219 \\
(4) Net count rate (cts s$^{-1}$)\tablenotemark{d} & 310 & 1838 & 346 & 2695 \\
(5) Stellar count rate (cts s$^{-1}$)\tablenotemark{d} & 14 & 61 & 3.1 & 12 \\
(6) Predicted count rate (cts s$^{-1}$)\tablenotemark{e} & 243 & 2031 & 279 & 3532 \\
(7) Predicted UV count rate (cts s$^{-1}$)\tablenotemark{f}& 78 & 1591 & 50 & 2929 \\
(8) (Net {\it FOC})/(UV+optical) & 1.28 & 0.90 & 1.24 & 0.76 \\
(9) (Stellar {\it FOC})/UV & 0.18 & 0.04 & 0.06 & 0.004\\
\tableline
\end{tabular}
\tablenotetext{a}{Measured count rate through an
artificial $10\times20\arcsec$ oval aperture (A+2B in Figures 4 \& 5).}
\tablenotetext{b}{Assumed dark count rate through an
artificial $10\times20\arcsec$ oval aperture 
(A+2B in Figures 4 \& 5), assuming nominal {\it FOC} dark current.}
\tablenotetext{c}{Assumed sky count rate through an
artificial $10\times20\arcsec$
oval aperture (A+2B in Figures 4 \& 5), assuming average
sky template spectrum.}
\tablenotetext{d}{Measured count rate for stars in the
artificial $10\times20\arcsec$
oval aperture (A+2B in Figures 4 \& 5) and
brighter than the limiting magnitudes.}
\tablenotetext{e}{Predicted count rate through a 
$10\times20\arcsec$ oval aperture, including red leak.}
\tablenotetext{f}{Predicted UV count rate through a 
$10\times20\arcsec$ oval aperture ($\lambda < 2400$~\.{A}
in F175W and $\lambda < 3400$~\.{A} in F275W).}
}

\end{table}

\end{document}